Marciani, G., Ronchitelli, A., Arrighi, S., Badino, F., Bortolini, E., Boscato, P., Boschin, F., Crezzini, J., Delpiano, D., Falcucci, A., Figus, C., Lugli, F., Oxilia, G., Romandini, M., Riel-Salvatore, J., Negrino, F., Peresani, M., Spinapolice, E.E., Moroni, A., Benazzi, S.

**Lithic technocomplexes in Italy from 50 to 39 thousand years BP: an overview of lithic technological changes across the Middle-Upper Palaeolithic boundary.**

# Lithic techno-complexes in Italy from 50 to 39 thousand years BP: An overview of lithic technological changes across the Middle-Upper Palaeolithic boundary

Giulia Marciani [a,b,*], Annamaria Ronchitelli [b], Simona Arrighi [a,b], Federica Badino [a,c], Eugenio Bortolini [a], Paolo Boscato [b], Francesco Boschin [b], Jacopo Crezzini [b], Davide Delpiano [d], Armando Falcucci [e], Carla Figus [a], Federico Lugli [a], Gregorio Oxilia [a], Matteo Romandini [a], Julien Riel-Salvatore [f], Fabio Negrino [g], Marco Peresani [c], Enza Elena Spinapolice [h], Adriana Moroni [b], Stefano Benazzi [a]

[a] *Università di Bologna, Dipartimento di Beni Culturali, Via degli Ariani 1, 48121, Ravenna, Italy*
[b] *Dipartimento di Scienze Fisiche, della Terra e dell'Ambiente, U. R. Preistoria e Antropologia, Università di Siena, Via Laterina 8, 53100, Siena, Italy*
[c] *C.N.R. - Istituto di Geologia Ambientale e Geoingegneria, 20126, Milano, Italy*
[d] *Dipartimento di Studi Umanistici, Sezione di Scienze Preistoriche e Antropologiche, Università di Ferrara, Corso Ercole I d'Este 32, 44100, Ferrara, Italy*
[e] *Department of Early Prehistory and Quaternary Ecology, Tübingen University.SchlossHohentübingen, D-72070, Tübingen, Germany*
[f] *Dipartimento di Antichità, Filosofia, Storia, Università degli Studi di Genova, Via Balbi 2, 16126, Genova, Italy*
[g] *Département d'anthropologie, Université de Montréal, 2900 Boulevard Edouard-Montpetit, Montréal, QC, H3T 1J4, Canada*
[h] *Dipartimento di Scienze dell'Antichità, Università degli Studi di Roma "La Sapienza", Piazzale Aldo Moro 5, 00185, Roma, Italy*

ABSTRACT

Defining the processes involved in the technical/cultural shifts from the Late Middle to the Early Upper Palaeolithic in Europe (~50-39 thousand years BP) is one of the most important tasks facing prehistoric studies. Apart from the technological diversity generally recognised as belonging to the latter part of the Middle Palaeolithic, some assemblages showing original technological traditions (i.e. Initial Upper Palaeolithic: Bohunician, Bachokirian; so called transitional industries: Châtelperronian, Szeletian, Lincombian-Ranisian-Jerzmanowician, Uluzzian; Early Upper Palaeolithic: Protoaurignacian, Early Aurignacian) first appear during this interval. Explaining such technological changes is a crucial step in order to understand if they were the result of the arrival of new populations, the result of parallel evolution, or of long-term processes of cultural and biological exchanges. In this debate Italy plays a pivotal role, due to its geographical position between eastern and western Mediterranean Europe as well as to it being the location of several sites showing Late Mousterian, Uluzzian and Protoaurignacian evidence distributed across the Peninsula. Our study aims to provide a synthesis of the available lithic evidence from this key area through a review of the evidence collected from a number of reference sites. The main technical features of the Late Mousterian, the Uluzzian and the Protoaurignacian traditions are examined from a diachronic and spatial perspective. Our overview allows the identification of major differences in the technological behaviour of these populations, making it possible to propose a number of specific working hypotheses on the basis of which further studies can be carried out. This study presents a detailed comparative study of the whole corpus of the lithic production strategies documented during this interval, and crucial element thus emerge: 1. In the Late Mousterian tools were manufactured with great attention being paid to the production phases and with great investment in inizializing and managing core convexities; 2. In contrast, Uluzzian lithic production proceeded with less careful management of the first phases of debitage, mainly obtaining tool morphologies by retouching. 3. In the Protoaurignacian the production is carefully organized and aimed at obtaining laminar blanks (mainly bladelets) usually marginally retouched. These data are of primary importance in order to assess the nature of the "transition" phenomenon in Italy, thus contributing to the larger debate about the disappearance of Neandertals and the arrival of early Modern Humans in Europe.

* Corresponding author. Università di Bologna, Dipartimento di Beni Culturali, Via degli Ariani 1, 48121, Ravenna, Italy.

*E-mail addresses:* giulia.marciani@unibo.it (G. Marciani); annamaria.ronchitelli@unisi.it (A. Ronchitelli); simona.arrighi@unibo.it (S. Arrighi); federica.badino@unibo.it (F. Badino); eugenio.bortolini2@unibo.it (E. Bortolini); paolo.boscato@unisi.it (P. Boscato); fboschin@hotmail.com (F. Boschin); jacopocrezzini@gmail.com (J. Crezzini); davide.delpiano@unife.it (D. Delpiano); armando.falcucci@ifu.uni-tuebingen.de (A. Falcucci); carla.figus3@unibo.it (C. Figus); federico.lugli6@unibo.it (F. Lugli); gregorio.oxilia3@unibo.it (G. Oxilia); matteo.romandini@unibo.it (M. Romandini); julien.riel-salvatore@umontreal.ca (J. Riel-Salvatore); fabio.negrino@unige.it (F. Negrino); marco.peresani@unife.it (M. Peresani); enzaelena.spinapolice@uniroma1.it (E.E. Spinapolice); adriana.moroni@unisi.it (A. Moroni); stefano.benazzi@unibo.it (S. Benazzi)

## 1. Introduction

Between 50-39 ka cal BP, Western Eurasia was the scene of one of the most debated events in prehistory: the demise of the autochthonous Neandertal populations and their replacement by Modern Humans (hereafter MHs). Along with this crucial biological turnover, significant techno-cultural changes took place among Palaeolithic hunter-gatherer societies, notably the introduction of novel lithic production techniques, of new bone and lithic tool types, as well as of the systematic use of ornamental objects and colouring substances (Mellars, 1989; Bar-Yosef, 2002).

Understanding the dynamics that pushed this complex phenomenon requires an in-depth knowledge of the biological and cultural processes that drove it, which reveals the potential for adaptation and innovation amongst both late Neandertals and early MHs (Bar-Yosef, 2002; Hardy et al., 2008; d'Errico et al., 2012; Hublin, 2012, 2015; Villa and Roebroeks, 2014; Fu et al., 2014; Higham et al., 2014; Hershkovitz et al., 2015; Pagani et al., 2015; Posth et al., 2016; d'Errico and Colagè, 2018). The period of interest falls in the middle of Marine Isotope Stage 3 (MIS 3: 60–30 ka cal BP) and was climatically unstable, with temperate phases interrupted by cold and often arid episodes in southern Europe known as Heinrich Event 5 (49-47 ka) and Heinrich Event 4 (40.2-38.3 ka) (Sánchez Goñi et al., 2008; Müller et al., 2011; Blockley et al., 2012). Further, HE 4 closely followed the volcanic event known as the Campanian Ignimbrite (40Ar/39Ar age: 39.85 ± 0.14 ka; Giaccio et al., 2017 and references therein).

In this context, both Neandertals and MHs had to develop, renew and update their ability to exploit resources in an environment characterized by highly diverse geomorphological and latitudinal conditions that often rapidly fluctuated as a result of climate change (Davies and Gollop, 2003; Davies et al., 2003, 2015; Stewart et al., 2003; Van Andel et al., 2003; Lowe et al., 2012).

As for the 50-39 ky cal BP interval, Hublin (2015) tentatively proposed the following four-part division of the cultural entities of Europe during MIS 3: 1. The Middle Palaeolithic techno-complexes; 2. The Initial Upper Palaeolithic techno-complexes (Emirian, Bohunician, Bachokirian) currently limiteded to eastern and central Europe; 3. The so-called "transitional" techno-complexes: the Châtelperronian and the Uluzzian in Western/Mediterranean Europe, the Neronian in southern France, the Lincombian-Ranisian-Jerzmanowician in Northern Europe and the Szeletian in Central-Eastern Europe; 4. The Upper Palaeolithic techno-complexes, namely the Protoaurignacian, Early Aurignacian, and Aurignacian techno-complexes.

The problem is that the four groups of techno-complexes almost completely overlap chronologically (e.g.Douka et al., 2014; Higham et al., 2014) and that, except for the Middle Paleolithic which is so far associated with Neanderthals (e.g. Schmitz et al., 2002; Lalueza-Foxet al., 2005 but cf. Harvati et al., 2019), very few of the assemblages composing them are associated with diagnostic human remains (e.g. Benazzi et al., 2011; Hublin, 2015; Gravina et al., 2018).

This, of course, raises the thorny question of whether one can extrapolate the biological identity of the makers (Neandertals or MHs) of a given assemblage based on an association at other sites, which on the strength of current evidence seems unwarranted (Hublin, 2015; Kuhn, 2018; Slimak, 2018).

Italy plays a pivotal role in this ongoing debate due to both its geographical position between eastern and western Mediterranean Europe, as well as to its vast ecological diversity. Numerous Late Mousterian, Uluzzian and Protoaurignacian sites, distributed across the Peninsula, provide us with the empirical basis for studying continuities and discontinuities in this mosaic of local traditions and new technological trends.

In Italy, the Mousterian is attributed to the Neandertals based on the association between Mousterian lithic assemblages and Neandertals fossils (Palma di Cesnola, 1996) at Buca del Tasso (Cotrozzi et al., 1985), Grotta delle Fate (Giacobini et al., 1984), Grotta Fumane (Benazzi et al., 2014), Riparo Tagliente (Arnaud et al., 2016), Grotta Nadale (Arnaud et al., 2017), Grotta Breuil (Manzi and Passarello, 1995Manzi and Passarello, 1995), Grotta del Fossellone (Mallegni, 1992), Grotta Guattari (Arnaud et al., 2015), Riparo del Molare (Mallegni and Ronchitelli, 1987), Grotta del Cavallo (Messeri and Palma di Cesnola, 1976; Fabbri et al., 2016), Grotta del Bambino (Blanc, 1962a, 1962b), Grotta Taddeo (Benazzi et al., 2011). The Late Mousterian is mainly characterised by the use of Levallois, discoid and unidirectional volumetric debitage, with a preference for the production of elongated blanks in its latest stages (e.g. Peresani, 2012; Gennai, 2016; Carmignani, 2017; Marciani, 2018). Sporadic use of ornaments (Romandini et al., 2014) and bone tools is documented (Jéquier et al., 2012; Romandiniet al., 2015).

The Uluzzian is currently considered to bea product of MHs (Benazzi et al., 2011; Moroni et al., 2013, 2018a; for an opposing view see Zilhão et al., 2015; Villa et al., 2018) mainly due to the two decidous teeth of Grotta del Cavallo. Its hallmarks are the significant use of the bipolar technique, the presence of lunates and the abundance of end-scrapers (Palma di Cesnola, 1989; Riel-Salvatore, 2009; Moroni et al., 2018a).

The Protoaurignacian is attibuted to MHs, as it has been confirmed by the two MH incisors retrieved at Riparo Bombrini and Grotta di Fumane (Benazzi et al., 2015). The main features of this techno-complex are cores meant for the recurrent production of blades and bladelets, as well as the use of marginally backed bladelets (Falcucci et al., 2018a, 2018b; Negrino and Riel-Salvatore, 2018; Riel-Salvatore and Negrino, 2018a). Both the Uluzzian and the Protoaurignacian are characterised by the occurrence of ornaments on marine shells, of bone points and of colouring substances, though there is some regional differention in their distribution across the Peninsula (Stiner, 1999; d'Errico et al., 2012; Tejero and Grimaldi, 2015 and see Arrighi et al., in this special issue).

The foremost aim of this paper is to give an updated synthesis of the lithic assemblages occurring during the "MP-UP transition" in Italy. We want to evaluate chronological changes involved, based on a critical review of the lithic techno-complexes in Italy found in layers from key stratified reference sites reliably dated to MIS3. This study intends to lay the foundations for the research already underway and to highlight gaps in our knowledge that remain to be filled by future work.

### *1.1. Geographical distribution, stratigraphies and chronology*

Sites are not evenly distributed, and several areas, such as the Po plain, the Apennine chain, the Adriatic coastal belt and the main islands, appear at the moment devoid of human occupation during MIS3, possibly due to the late Pleistocene-Holocene geomorphic evolution of these regions (Antonioli and Vai, 2004). Conversely, there are some regions/districts of variable size with clusters of sites (Fig. 1). This pattern, which remains unvaried from the Late Mousterian to the Uluzzian until the Protoaurignacian independently of the number of sites involved, results from a combination of causes such as: 1) the occurrence of climatically and environmentally favourable niches; 2) the possibility that several sites were eroded or buried because of geological events; 3) the Late-glacial and Holocene marine transgression which submerged coastal sites (Antonioli, 2012); and 4) the uneven spatial distribution and development of field investigation and research.

The presence of the Late Mousterian, the Uluzzian and the Protoaurignacian has been ascertained in many caves and rock shelters, four of which (Grotta di Fumane, Grotta La Fabbrica, Grotta di Castelcivita, Grotta della Cala) have yielded sequences containing all three techno-complexes. In these sites, the sequence (from the bottom) is consistently Mousterian – Uluzzian – Protoaurignacian, according to both the stratigraphic position and the chronological data (Fig. 2a). Otherwise, in the absence of the Uluzzian, the sequence is Mousterian – Protoaurignacian (Mochi and Bombrini) or Mousterian – Uluzzian (Grotta del Cavallo, Riparo del Broion) (Fig. 2b). No evidence of interstratification has ever been found in Italy. Importantly, a stratigraphic discontinuity, usually represented by volcanic layers, erosional events or sedimentary hiatuses, recurs between the Mousterian and the overlying layers in southern Italy, suggesting a break in the human occupation of these sites (Fumanal, 1997; Peresani et al., 2014; Moroni et al., 2018a; Zanchetta et al., 2018) and in Liguria at Riparo Mochi (Grimaldi et al., 2014).

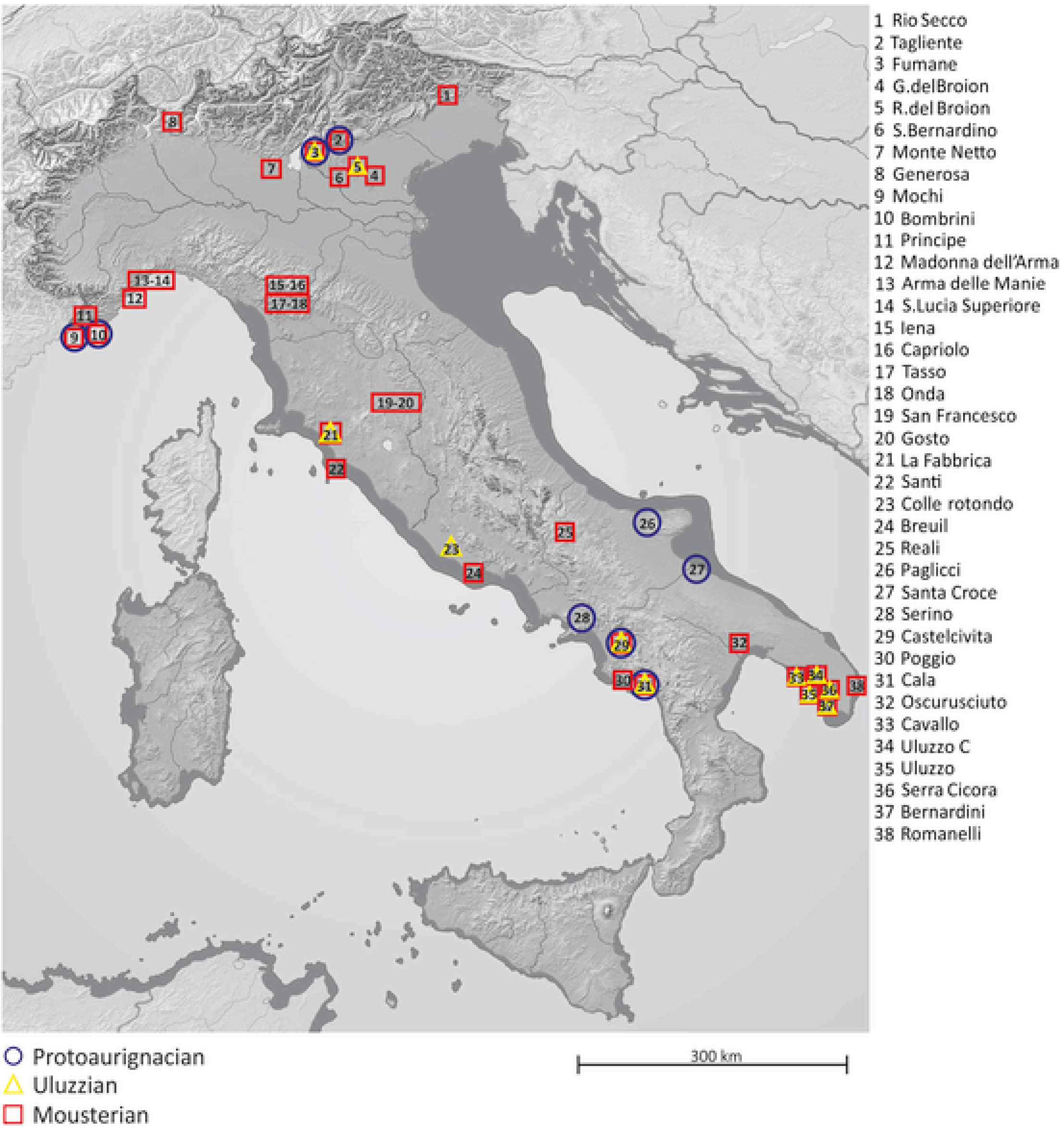

**Fig. 1.** Location of the Italian key sites with MIS3 human occupations. The Italian Peninsula shows a sea level of 70 m below the present-day coastline, based on the global sea-level curve (Benjamin et al., 2017) but lacking the estimation of post-MIS3 sedimentary thickness and eustatic magnitude (sketch map courtesy by S. Ricci, University of Siena).

In the Salento region, several sites with Uluzzian occupations (Grotta del Cavallo, Grotta di Uluzzo, Grotta di Uluzzo C, Grotta di Serra Cicora, Grotta Mario Bernardini and Grotta delle Veneri) are clustered within an area of a few km$^2$. Among these, Grotta del Cavallo remains up till now the type site, with optimal condition for studying the Uluzzian in its chrono-cultural development. Here three main phases have been identified: the archaic Uluzzian (layer EIII), the evolved Uluzzian (layers EII-I) and the late (or upper) Uluzzian (layer D) (Palma di Cesnola, 1993). The archaic Uluzzian is also recorded at Bernardini (layer AIV); while the Upper Uluzzian has been found at Uluzzo (layer N), Uluzzo C (layers D-C) and Bernardini (layer AII-I). According to Palma di Cesnola, the very final Uluzzian is absent at Grotta del Cavallo. It has, however, been identified in the nearby cave of Serra Cicora (horizon D of layer B) where it is followed by the so-called "Uluzzo-Aurignaziano" (A, B and C of layer B) (Palma di Cesnola, 1993). Outside Apulia the archaic phase is not recorded: Grotta della Cala and Grotta di Castelcivita (Campania) yielded evidence of the final and evolved phases respectively (Benini et al., 1997; Gambassini, 1997). Grotta della Fabbrica (Tuscany) contains a late or final Uluzzian occupation (Dini and Conforti, 2011; Dini and Tozzi, 2012; Villa et al., 2018), and Grotta di Fumane and Riparo del Broion have been attributed to the evolved Uluzzian (Peresani et al., 2019).

The boundary between the Middle and the Upper Palaeolithic falls in a time span close to the limit of radiocarbon dating capability. However, over the last years, advancements in dating methods, such as $^{14}C$, TL and OSL, have significantly improved the reliability of dates on charcoal (ABOx) bone (ultrafiltration), shells and sediments (Higham, 2011), leading to refine the timing of the Middle to Upper Palaeolithic shift in Europe. According to Higham and colleagues (2014), MHs and indigenous populations coexisted in Europe for at least 5400-

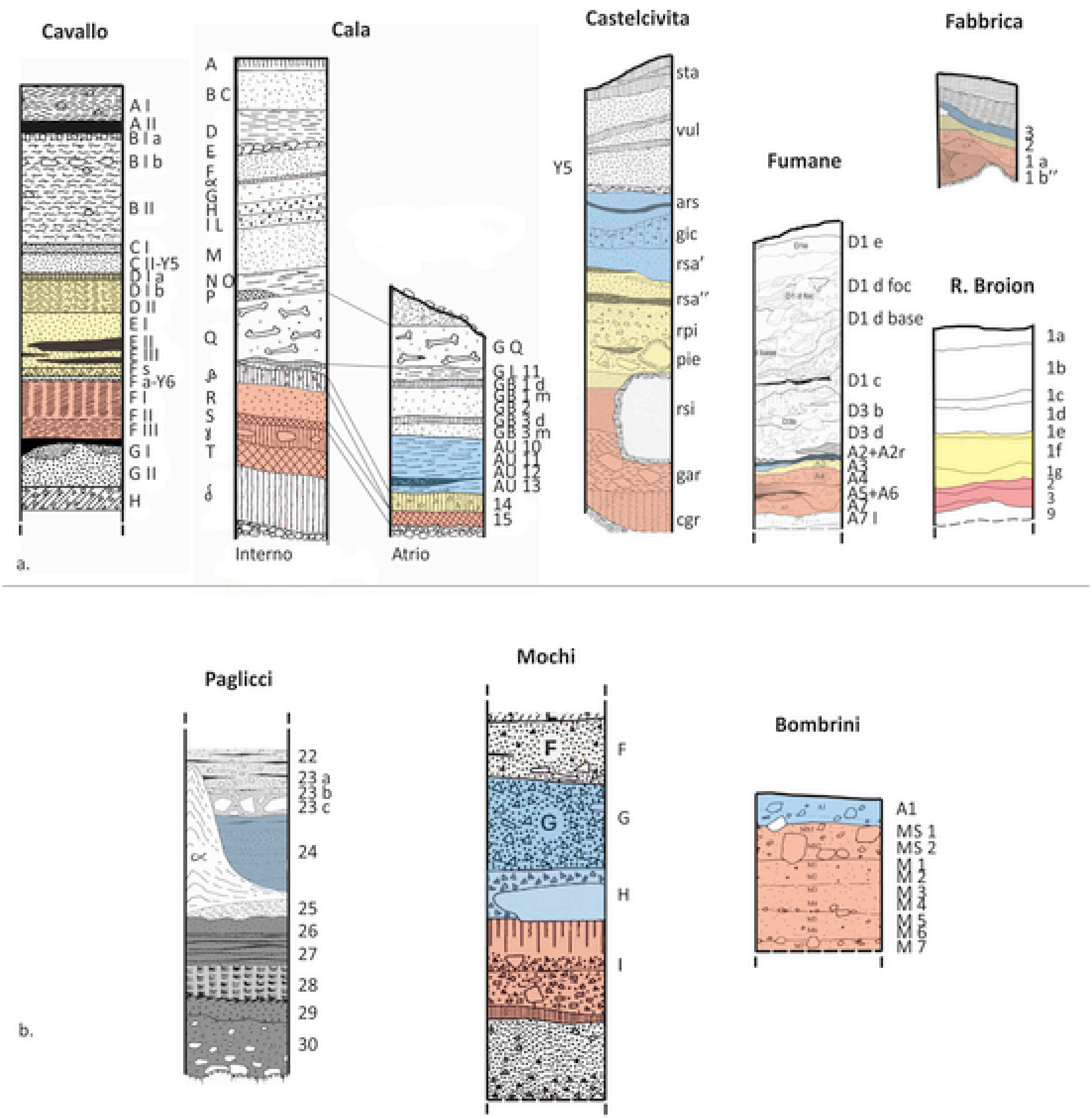

**Fig. 2.** Key stratigraphic sequences. a. Mousterian, Uluzzian and Protoaurignacian sequences (Cavallo modified from Palma di Cesnola, 1964; Cala, Castelcivita modified from Gambassini, 1982; Fumane modified from Tagliacozzo et al., 2013; R. Broion modified from Peresani et al., 2019; Fabbrica modified from Villa et al., 2018); b. Mousterian-Protoaurignacian sequences (Paglicci modified from Palma di Cesnola, 1992; Mochi modified from de Lumley, 1969; Bombrini modified from Riel-Salvatore and Negrino, 2018).

2600 years (probability: 95.4%), since Neandertals definitively disappeared from western and central Europe about 41,030-39,260 years cal BP.

Due to the intense activity of Italian volcanoes, Mediterranean Palaeolithic stratigraphic sequences often contain tephra layers, which represent important chronological markers also functioning as isochrons on a large geographical scale. At Grotta del Cavallo the whole Uluzzian package is sandwiched between two tephra layers (Fa and CII) - the Y-6 green tuff of Pantelleria and the Y-5 Campanian Ignimbrite (CI) dated at 45.5 ± 1.0 ka and 39.85 ± 0.14 ka respectively (Zanchetta et al., 2018). The latter of which embodies, in fact, the very chronological limits of the Uluzzian techno-complex in southern Italy (Table 1 SM, Fig. 3). These limits are entirely consistent with the radiocarbon chronological model obtained from the same site (Benazzi et al., 2011; Douka et al., 2014). Tephra layers attributed to the CI were found to seal the Protoaurignacian at the Campanian sites of Castelcivita (Gambassini, 1997) and Serino (Accorsi et al., 1979) and overlie the Uluzzian context at Klissoura Cave (Greece) (Koumouzelis et al., 2001; Stiner et al., 2007) and a possible Uluzzian evidence at Crvena Stijena (Montenegro) (Mihailović and Whallon, 2017; Morin and Soulier, 2017).

Based on the radiometric evidence, from 50 to 39 ka cal BP the Italian territory is characterised by the synchronous occurrence of various cultural entities both in the North and in the South (Fig. 4). Although Grotta della Cala is one of the key sequences of the MP to UP transition in Italy (and for this reason has been included in Table 1 SM; Fig. 1), radiocarbon dates obtained from this site for the Uluzzian and Protoaurignacian levels must be considered unreliable as they are much too young relative to those for comparable assemblages at other sites. This issue is currently under investigation and a new dating project, including OSL dating, has been launched at this cave by the Oxford University Lab. As research currently stands, the oldest evidence of the Uluzzian is recorded in southern Italy (Grotta del Cavallo) (Benazzi et al., 2011; Zanchetta et al., 2018), whereas the oldest dates relating to the Protoaurignacian have been found in northern Italy (Mochi and Fumane) (Douka et al., 2012, 2014) (Table 1 SM; Fig. 3). Assuming that exogenous populations of MHs introduced both the Uluzzian and the Protoaurignacian, radiometric data seem to indicate different migration routes. If the available dates converge to suggest a quasi-linear north-to-south diffusion of the Protoaurignacian (Palma di Cesnola, 2004b), a different and more complex model seems to apply to

**Table 1**
Raw material, concept of debitage and target objectives of the Italian key site pertaining to the Mousterian, Uluzzian and Protoaurignacian techno-complex.

| Site\layer | Main raw material | Secondary raw material | Source of raw material | MainTypeofrawmaterial | SecondaryTypeofrawmaterial | Main concept of debitage | Secondary concept of debitage | Production structure | Main objective of debitage | Secondary objective of debitage | Main retouched tools | Secondary retouched tools | References |
|---|---|---|---|---|---|---|---|---|---|---|---|---|---|
| **Mousterian** | | | | | | | | | | | | | |
| Rio secco, 5-7-8 | Flint | | Local, exogenous | Nodules | Pebbles | Unipolar Levallois | Centripetal Levallois, discoid | Integrated | Flakes | | Scraper | | Peresani et (2014) |
| San Bernardino, II | Flint | | Local, exogenous | Blocks | Nodules | Unipolar Levallois | Centripetal Levallois | Integrated | Flakes | | | | (Peresani, Picin et al. 2013; Peres al., 2015) |
| Broion grotta, H-N1 | Flint | | Local, exogenous | | | Unipolar Levallois | Centripetal Levallois | Integrated | Flakes | | | | (Peresani a Porraz, 20 Peresani, 2 |
| Fumane, A4-A6 | Flint | Limestone | Local | Nodules | Pebbles | Unipolar Levallois | Centripetal Levallois, blade volumetric concept | Integrated, Additional | Blades | Elongated flakes | Scraper | | (Peresani, 2011; 2012 Gennai, 20 Peresani et 2013, 2016 |
| Fumane, A8-A9 | Flint | | Local | Blocks | Flakes | Discoid | | Integrated | Flakes | | | Backed pieces | (Peresani, 2012; Genn 2016; Delp and Peresa 2017; Delp et al., 201 Delpiano e 2019) |
| Fumane, A 10 | Flint | | Local | Nodules | Pebbles | Unipolar Levallois | Centripetal Levallois, discoid | Integrated | Flakes | Elongated flakes | Scraper | Denticulate | (Peresani, 2012; Genn 2016) |
| Monte Netto | Flint | Quartztite | Local | | | Levallois | | Integrated | Blades | Flakes | Scraper | | Delpiano e (in press) |
| Generosa, 2, 11,12 | Radiolarite | | Local, exogenous | | | Unipolar Levallois | Discoid | Integrated | Flakes | | Denticulate | | Bona et al. (2007) |
| Mochi, I | Flint | Quartzitic sandstone | Local | Pebbles | | Unipolar Levallois | Centripetal Levallois, discoid, unipolare | Integrated, Additional | Blades | Flakes, elongated flakes | Scraper | | (Grimaldi a Santaniello 2014; Ross Notter et a 2017; Negr and Riel-Salvatore, |
| Bombrini, IV | Flint | Quartzitic sandstone | Local | | | Discoid | Centripetal Levallois | Integrated | Flakes | | Scraper | Denticulate | (Rossoni-N et al., 2017 Negrino an Riel-Salvat 2018; Riel-Salvatore a Negrino, 2 |

**Table 1** (*Continued*)

| Site\layer | Main raw material | Secondary raw material | Source of raw material | MainTypeofrawmaterial | SecondaryTypeofrawmaterial | Main concept of debitage | Secondary concept of debitage | Production structure | Main objective of debitage | Secondary objective of debitage | Main retouched tools | Secondary retouched tools | References |
|---|---|---|---|---|---|---|---|---|---|---|---|---|---|
| Principe, A1 | Limestone | Quartzitic sandstone | Local | | | Levallois | Discoid | Integrated | Blades | Flakes, elongated flakes | Scraper | | (De Lumley al., 2008; Negrino an Tozzi, 2008 Rossoni-No et al., 2017 |
| Madonna dell'Arma, I-II | Quartzitic sandstone | | Local | | | Unipolar Levallois | Centripetal Levallois, discoid, laminar | Integrated, Additional | Blades | Flakes | Scraper | Notch | (Cauche, 2( Rossoni-No et al., 2017 |
| Arma delle Manie, II | Dolomite | Quartzitic sandstone | Local | Blocks | Pebbles | Discoid | Levallois, centripetal | Integrated, Additional | Flakes | | Scraper | | Leger (201 |
| Gosto D, C | | | | Pebbles | | Levallois | | Integrated | Flakes | | Scraper | | (Tozzi, 197 Casini, 201 |
| Santi | Radiolarite | Siliceous limestone | Local | Pebbles | | Unipolar Levallois | Volumetric unidirectionl production with management of striking platform | Integrated, Additional | Flakes | Elongated flakes | | | (Moroni et 2008, Morc al., 2018) |
| La Fabbrica, 1 | Radiolarite | Flint | Local | Pebbles | | Centripetal Levallois | unidirectional, bidirectional or centripetal debitage with any special preparation of the debitage surface or core shaping | Integrated, Additional | Flakes | Elongated flakes | Scraper | Denticulate | (Dini et al. 2007; Dini Conforti, 2 Dini and T( 2012; Villa al., 2018) |
| Breuil, 3-6 | Flint | | | Pebbles | | Unidirectional | Bidirectional, pseudo prismatic, Levallois | Additional, Integrated | Blades | Flakes | Scraper | Notch, denticulate | (Lemorini, 2 Grimaldi ai Spinapolice 2010; Grim and Santan 2014) |
| Breuil, 7-8 | Flint | | | Pebbles | | Unidirectional | Bidirectional, pseudo prismatic cores | Additional | Flakes | | Scraper | Notch, denticulate | (Lemorini, 2 Grimaldi ai Spinapolice 2010; Grim and Santan 2014) |
| Reali, 2-5 | Flint | Flint | Local | Slabs | | Orthogonal plans | Unipolar Levallois, centripetal Levallois, discoid | Additional, integrated | Flakes | Blades | Scraper | Notch | (Arzarello ( 2004; Rufo 2008; Peret 2012) |
| Castelcivita, rsi-cgr-gar | | | | Blocks | | Unipolar Levallois | Centripetal Levallois | Integrated | Blades | Elongated flakes, flakes | | | Study ongoi ERC; (Gambassir 1997) |
| Cala, R; 15 | | | | | | Preferencial Levallois | | Integrated | Flakes | | Scraper | | Caramia (2 |

**Table 1** (*Continued*)

| Site\layer | Main raw material | Secondary raw material | Source of raw material | MainTypeofrawmaterial | SecondaryTypeofrawmaterial | Main concept of debitage | Secondary concept of debitage | Production structure | Main objective of debitage | Secondary objective of debitage | Main retouched tools | Secondary retouched tools | References |
|---|---|---|---|---|---|---|---|---|---|---|---|---|---|
| Poggio, 9-10 | Flint | Flint | Local, exogenous | Pebbles | | Unipolar Levallois | Centripetal Levallois | Integrated | Blades | Flakes | Denticulate | Scraper | (Caramia a Gambassin 2006; Bosc al., 2009) |
| Oscurusciuto, 1-4 | Radiolarite | Flint, siliceous limestone | Local | Pebbles | | Unipolar Levallois | Centripetal Levallois, bladelets volumetric production | Integrated, additional | Blades | Elongated flakes, flakes | Scraper | | (Boscato et 2011; Ronc et al., 2011 Ranaldo et 2017b; Mar 2018) |
| Romanelli G | Limestone | Siliceous limestone | Local | Pebbles | | Levallois | Opportunistic | Integrated, additional | Flakes | | Notch | Scraper | (Piperno, 1 Spinapolice 2008) |
| Cavallo, FII-FIIIa | Siliceous limestone | Limestone | Local | Blocks | Slabs | Discoid | Kombewa, orthogonal plans | Integrated, additional | Flakes | | Scraper | Denticulate | (Carmignan 2010, 2011 |
| Cavallo, FIIIb-FIIIc | Siliceous limestone | Limestone | Local | Blocks | Slabs | Centripetal Levallois | Unipolar parallal plans, orthogonal plans | Integrated, additional | Blades | Flakes | Scraper | Denticulate | (Carmignan 2010, 2011 |
| Uluzzo C; G | Limestone | Siliceous limestone | Local | Slabs | Blocks | Levallois | | integrated | Flakes | | Scraper | Point | (Borzatti v Löwensterr 1966; Spinapolice 2008, 2012 |
| Bernardini, B1 | Siliceous limestone | Limestone | Local | Slabs | | Discoid | | Integrated | Flakes | | Denticulate | Notche | (Borzatti v Löwensterr 1970; Spinapolice 2008; Carmignan 2011; Romagnoli, 2012) |
| Bernardini, B3-B4 | Siliceous limestone | Limestone | Local | Slabs | | Centripetal Levallois | Volumetric debitage blade | Integrated, additional | Blades | Flakes | Denticulate | Notch | (Borzatti v Löwensterr 1970; Borz von Löwen 1965Spina 2008; Carmignan 2011; Romagnoli, 2012) |
| **Uluzzian** | | | | | | | | | | | | | |

**Table 1** (*Continued*)

| Site\layer | Main raw material | Secondary raw material | Source of raw material | MainTypeofrawmaterial | SecondaryTypeofrawmaterial | Main concept of debitage | Secondary concept of debitage | Production structure | Main objective of debitage | Secondary objective of debitage | Main retouched tools | Secondary retouched tools | References |
|---|---|---|---|---|---|---|---|---|---|---|---|---|---|
| Cavallo, E III | Limestone | Flint | Local | Slabs | Pebbles | Bipolar production | Direct freehand percussion, debitage is very simple as it encompasses none or only a minimal preparation of the volume to be flaked. Striking platforms are generally natural. knapping is unifacial (both unidirectional and bidirectional) | Additional | Flaklets | Bladelets | End scraper | Side scraper, backed tool | (Palma di Cesnola, 19 1964; Morc al., 2018a) |
| Cavallo, EI- II | | | | | | | | | | | | | (Palma di Cesnola, 19 1964). Stud ongoing ERC |
| Cavallo, D | | | | | | | | | | | | | (Palma di Cesnola, 19 1964). Stud ongoing ERC |
| Uluzzo C | Flint | Siliceous limestone | | Slabs | Pebbles | Unidirectional debitage with no or few management of the convexities characterised by the use of the bipolar technique. | Unidirectional volumetric production where the striking platform and the lateral and distal convexities are managed, with a direct percussion technique. | Additional | Bladelets | Flaklets | Side scraper | End scraper | Study ongoi |
| **Cicora A** | | | | | | | | | | | | | |
| Bernardini A I-IV | | | | | | | | | | | | | |
| Cala, 14 | Flint | Radiolarite | Local | Pebbles | | Single percussion plan, unilateral debitage | Bipolar production | Additional | Long flakes | Blades | Scraper | Denticulate | Benini et a (1997) |

**Table 1** (*Continued*)

| Site\layer | Main raw material | Secondary raw material | Source of raw material | MainTypeofrawmaterial | SecondaryTypeofrawmaterial | Main concept of debitage | Secondary concept of debitage | Production structure | Main objective of debitage | Secondary objective of debitage | Main retouched tools | Secondary retouched tools | References |
|---|---|---|---|---|---|---|---|---|---|---|---|---|---|
| Castelcivita, rsa", rpi, pie, rsi | Flint | Limestone | Local | Slabs | Pebbles | Unidirectional debitage wich exploit a single debitage surface and hierarchization between the striking platform and debitage surface | Bipolar production | Additional | Flaklets | Bladelets | Scaled piece | Denticulate | Study ongoi ERC; (Gambassi 1997) |
| Colle rotondo | Flint | | | Pebbles | Cobbles | Unidirectional, bidirectional or multidirectional cores with parallel removals on one or more debitage surfaces and a cortical or natural surface platform or a platform formed by a single large scar or multiple previous scars | Bipolar production | Additional | Flakes | Bladelets | Scaled piece | Denticulate | Villa et al. (2018) |
| Fabbrica 2 | Radiolarite | | | | | Unidirectional cores with parallel removals on a single debitage surface or two adjacent debitage surfaces and a platform formed by a single scar or from previous scars on the opposed surface | Bipolar production | Additional | Flakes | Bladelets | Scaled piece | Scraper | (Dini and T 2012; Villa al., 2018) |

**Table 1** (*Continued*)

| Site\layer | Main raw material | Secondary raw material | Source of raw material | MainTypeofrawmaterial | SecondaryTypeofrawmaterial | Main concept of debitage | Secondary concept of debitage | Production structure | Main objective of debitage | Secondary objective of debitage | Main retouched tools | Secondary retouched tools | References |
|---|---|---|---|---|---|---|---|---|---|---|---|---|---|
| Broion riparo, 1f, 1g | Flint | | Local, exogenous | | | Unidirectional debitage exploiting the major axis through striking on a flat surface (the butt, a cortical side, a pre-existing sharp fracture, etc.) by bipolar technique | Unipolar production | Additional | Bladelets | Flakes | Backed piece | Scraper | Peresani et (2019) |
| Fumane, A3 | Flint | Limestone | Local | Nodules | Pebbles | Recurrent centripetal | Unipolar production-Laminar lamellar production | Additional | Flakes | Blades | Scraper | Denticulate | (Peresani e 2016, 2019 |
| **Protoaurignacian** | | | | | | | | | | | | | |
| Fumane, A1-A2 | Flint | | Local | Nodules | | Laminar and lamellar production. Platform cores | Multidirectional production | Integrated | Bladelets | Blades | Retouched bladelet | End scraper | (Falcucci e 2017, 2018 Falcucci an Peresani, 2 |
| Mochi G | Flint | | Exogenous, local | | | Laminar and lamellar production, prismatic unidirectional cores | | Integrated | Blades | Bladelets | Backed piece | Dufour bladelet | (Riel-Salva and Negrin 2009; Berto al., 2013; Grimaldi e 2014) |
| Bombrini A1-3 | Flint | | Exogenous, local | | | Laminar and lamellar production, prismatic unidirectional cores | Opportunistic production | Integrated | Bladelets | Flakes | Dufour bladelet | End scraper | (Riel-Salva 2007; Riel-Salvatore a Negrino, 2 2018b; Neg and Riel-Salvatore, |
| Fabbrica 3,4 | Flint | Quartz | Local, exogenous | Pebbles | Slabs | Laminar and lamellar production. Poliedric or prismatic cores | Opportunistic multidirectional production | Integrated | Flakes | Blades | Scraper | End scraper, retouched bladelet | Dini and T (2012) |
| Paglicci 24 | Flint | | Local | | | Laminar and lamellar production | | Integrated | Blades | Bladelets | Backed piece | End scraper, micropoint | Palma di C (2004b) |
| Serino | Flint | Radiolarite | Local, exogenous | Pebbles | | Bipolar production | Laminar and lamellar production | Integrated | Flakes | Blades, bladelets | Scraper | Truncation | Accorsi et (1979) |
| Cala AU 13-10 | Flint | Radiolarite | Local, exogenous | Pebbles | | Laminar and lamellar production | | Integrated | Flakes | Blades | End scraper | Scraper | Benini et a (1997) |

**Table 1** (*Continued*)

| Site\layer | Main raw material | Secondary raw material | Source of raw material | MainTypeofrawmaterial | SecondaryTypeofrawmaterial | Main concept of debitage | Secondary concept of debitage | Production structure | Main objective of debitage | Secondary objective of debitage | Main retouched tools | Secondary retouched tools | References |
|---|---|---|---|---|---|---|---|---|---|---|---|---|---|
| Castelcivita rsa' | Flint | Limestone | Local | Pebbles | | Laminar and lamellar production | | Integrated | Flakes | Blades, bladelets | Backed piece | End scraper, denticulate | Gambassini (1997) |
| Castelcivita ars, gic | Flint | Limestone | Local | Pebbles | | Laminar and lamellar production | | Integrated | Bladelets | Blades | Backed piece | End scraper, micropoint | Gambassini (1997) |

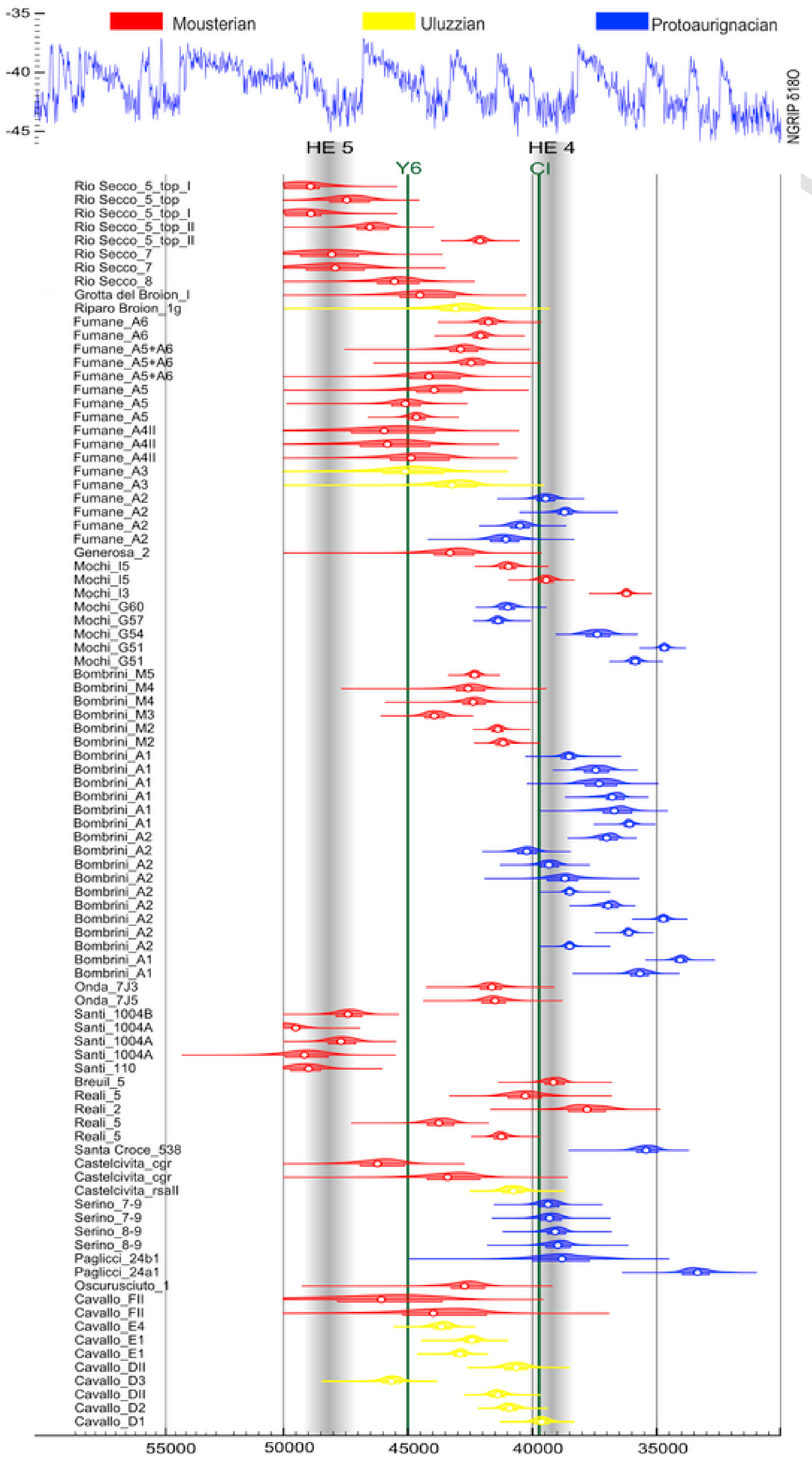

Mousterian
Uluzzian
Protoaurignacian
-35
-40
-45
NGRIP δ18O
HE 5
HE 4
Y6
CI
Rio Secco_5_top_I
Rio Secco_5_top
Rio Secco_5_top_I
Rio Secco_5_top_II
Rio Secco_5_top_II
Rio Secco_7
Rio Secco_7
Rio Secco_8
Grotta del Broion_I
Riparo Broion_1g
Fumane_A6
Fumane_A6
Fumane_A5+A6
Fumane_A5+A6
Fumane_A5+A6
Fumane_A5
Fumane_A5
Fumane_A5
Fumane_A4II
Fumane_A4II
Fumane_A4II
Fumane_A3
Fumane_A3
Fumane_A2
Fumane_A2
Fumane_A2
Fumane_A2
Generosa_2
Mochi_I5
Mochi_I5
Mochi_I3
Mochi_G60
Mochi_G57
Mochi_G54
Mochi_G51
Mochi_G51
Bombrini_M5
Bombrini_M4
Bombrini_M4
Bombrini_M3
Bombrini_M2
Bombrini_M2
Bombrini_A1
Bombrini_A1
Bombrini_A1
Bombrini_A1
Bombrini_A1
Bombrini_A1
Bombrini_A2
Bombrini_A2
Bombrini_A2
Bombrini_A2
Bombrini_A2
Bombrini_A2
Bombrini_A2
Bombrini_A2
Bombrini_A2
Bombrini_A1
Bombrini_A1
Onda_7J3
Onda_7J5
Santi_1004B
Santi_1004A
Santi_1004A
Santi_1004A
Santi_110
Breuil_5
Reali_5
Reali_2
Reali_5
Reali_5
Santa Croce_538
Castelcivita_cgr
Castelcivita_cgr
Castelcivita_rsaII
Serino_7-9
Serino_7-9
Serino_8-9
Serino_8-9
Paglicci_24b1
Paglicci_24a1
Oscurusciuto_1
Cavallo_FII
Cavallo_FII
Cavallo_E4
Cavallo_E1
Cavallo_E1
Cavallo_DII
Cavallo_D3
Cavallo_DII
Cavallo_D2
Cavallo_D1
55000
50000
45000
40000
35000

**Fig. 3.** 14C dates of Protoaurignacian, Uluzzian and Mousterian sites. OxCal v4.3.2 (Bronk Ramsey, 2017); r:5Int Cal 13; 68.2% atmospheric curve (Reimer et al., 2013). Raw dates and bibliographic references in Table 1 SM.

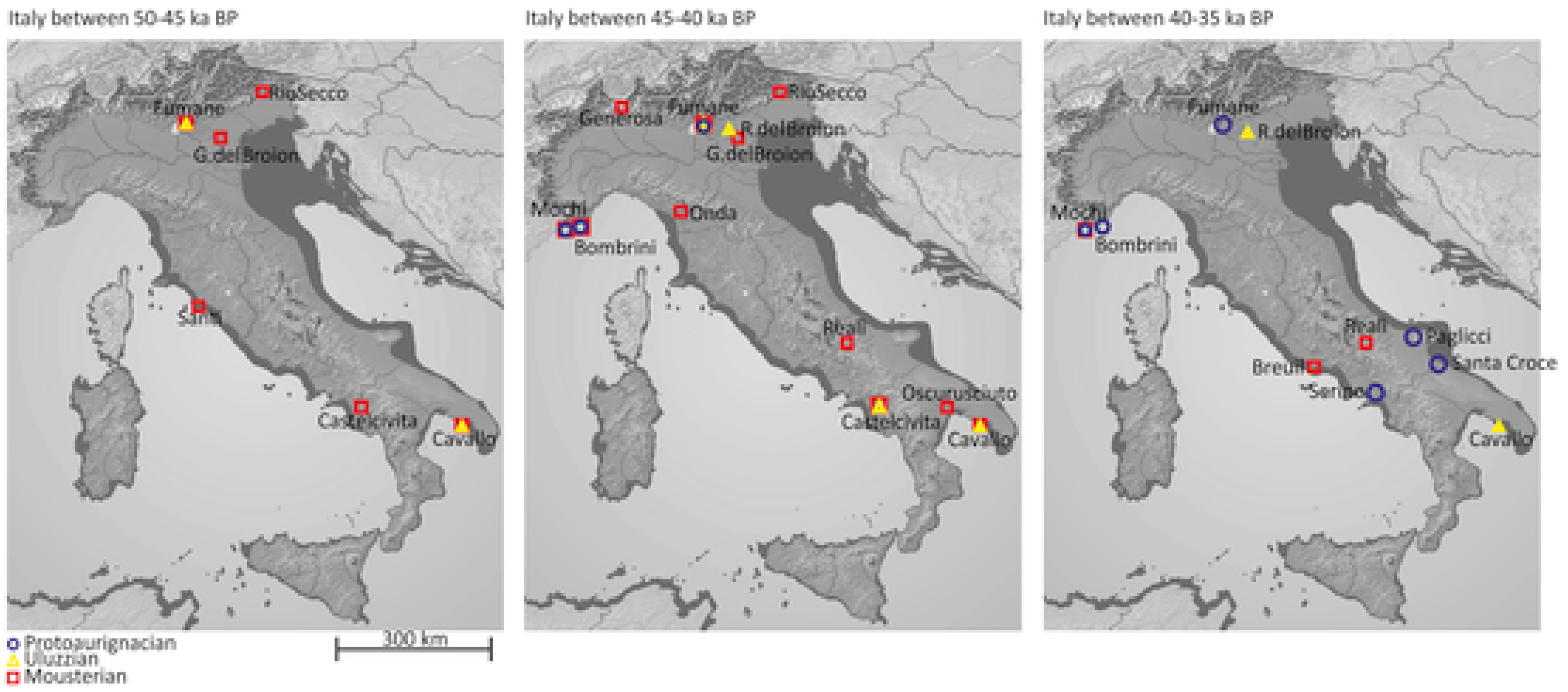

**Fig. 4.** Location of the Italian Mousterian, Uluzzian and Protoaurignacian key sites with MIS3 human occupations according to their 14C dates. Raw dates and bibliographic references in Table 1 SM.

the Uluzzian. Admitting the notion of an external provenance of this technocomplex (Moroni et al., 2013) (contrary to a local origin and development of it, see for instance Greenbaum et al., 2018), there is the possibility that the Uluzzian groups followed two different routes into Italy: the one through the Otranto channel up to the Ionian coast of the Salento and the other along the Balkanic coast of the Adriatic and then across the now-submerged Great Adriatic Plain up to the Colli Berici. The occurrence of several Uluzzian assemblages along the Tyrrhenian side as far north as Tuscany could be explained by Uluzzian groups migrating northwest from the core area in the Salento. This hypothesis is consistent with the more recent chronologies of the Campanian (Castelcivita, Cala) and the Tuscan (La Fabbrica) (Villa et al., 2018) assemblages.

## 2. Material and method

### *2.1. Analytic method*

In order to build a consistent reference section, we conducted a thorough review of the relevant scientific literature for the interval 50,000-39,000 years BP, limiting our geographical scope to Italy. The bibliographic references are based on main scientific publications such as journal articles, Master's and PhD theses, conference proceedings and other subject-specific publications in English, Italian, Spanish and French.

All the data was registered and standardised in an Access database. The criteria recorded to refer to the general characterisation of an assemblage and its location are: name of the site, type of site, geographical coordinates (when available), region, levels. For the chronology, we collected: MIS, laboratory code, date range, dating method, calibrated BP range (68,2%), indirect dating. Finally, for the lithic collection, which is the main object of this work, we recorded information about the structural conception of debitage (Boëda, 2013) and concept and method of debitage (Inizan et al., 1995): discoid (Boëda, 1993; Peresani, 2003); Levallois, preferential or recurrent unipolar, centripetal or convergent (Boëda, 1994); Kombewa (Owen, 1938); SSDA (Forestier, 1993); laminar and lamellar debitage (Boëda, 1990; Révillion and Tuffreau, 1994); and target product (flake, flakelets, blade, bladelets) (Inizan et al., 1995); then the types of most commonly used raw material (e.g. chert, jasper, limestone, siliceous limestone) and the type of initial block (pebbles, nodules, slabs or others).

For the Mousterian, we documented 29 assemblages drawn from 24 sites, most of them recently studied with the technological approach so almost all the data were updated and available. For the Uluzzian, we documented 13 assemblages from 11 archaeological sites. However, only five assemblages were published: Cavallo E III (Moroni et al., 2018a), Colle Rotondo (Villa et al., 2018), Fabbrica 2 (Villa et al., 2018), Riparo Broion 1f, 1g (Peresani et al., 2019), and Fumane A3 (Peresani et al., 2016, 2019). Castelcivita rsa", rpi, pie and rsi; Uluzzo C and Cavallo EII-I and D are currently under investigation so we present here unpublished data. For the Protoaurignacian, we documented assemblages from 9 sites; however technological data are currently available only for Fumane A1-A2 (Falcucci et al., 2017, 2018; Falcucci and Peresani, 2018), Mochi G (Bertola et al., 2013; Grimaldi et al., 2014) and Bombrini A1-3 (Riel-Salvatore, 2007; Negrino and Riel-Salvatore, 2018; Riel-Salvatore and Negrino, 2018a and b). It is clear that there is a disparity of studied assemblages across the three techno-complexes, which is the result of a combination of the amount of studied sites, the analytical protocol use, and the accessibility of raw data. For this reason, in this paper, we only present qualitative data. For a specific evaluation of quantitative data for each site/assemblage, it is necessary the access to raw data which is not available at the moment.

### *2.2. Vocabulary and epistemological basis*

In lithic studies, vocabulary is a problematic issue because it reflects distinct schools of thought and different approaches. Thus adopting a particular vocabulary requires the comprehension of the philosophical and epistemological world from which it derives. Consequently, choosing a particular terminology is both an epistemological choice and a philosophical statement.

One of our most challenging tasks was to standardise the terms that each scholar used to refer to each matter relating to lithic studies (i.e. technique, method and concept of debitage) in order to be able to compare assemblages studied by scholars with distinct backgrounds. In order to give here a comprehensive representation of this vocabulary relating to specific lithic production, we chose to interpret the available bibliography according to Boëda (2013).

The 'Boëda approach' is used by a restricted group of researchers and has been codified in the recent text "*Techno-logique & Technologie: Une paléo-histoire des objets lithiques tranchants*" (Boëda, 2013).

This volume admittedly articulates a novel framework that has been criticized by some as lacking in general descriptive potential and applicability (Frick and Herkert, 2014). However, because it allows for a systematic definition of each assemblage, we have adopted it here because it allows an evaluation not only of the technical features of each stone tool, but also the ideal template on which it was based and that characterises each techno-complex. Especially as concerns cores, it help to overcome the limitation imposed by classifications based solely on core morphology or scar directions and permits a complete comprehension of the volumetric and structural identity of each item.

Depending on the management of the block to be flaked, its volumetric and structural analysis, and the end-products, Boëda establishes a fundamental division between additional and integrated core types.

The additional core types include pieces where only part of the volume of the block is utilized as a core. That is, the core is made up of two independent parts: one is the active volume or the used portion, *alias* the core *sensu stricto*; the other is the passive volume, the block portion which is not necessary for the realisation of the objective. Thus, in an additional core it is possible to have two or more useful volumes (cores) in the same block, which means two or more series of blank can be knapped completely independently of each other. Among others, the types of debitage identified as additional are: debitage of orthogonal planes and opportunistic surface exploitation*, system par surface de débitage alterné* - SSDA (Forestier, 1993; Ashton et al., 1994), volumetric laminar production without management of convexities (Guilbaud and Carpentier, 1995; Boëda, 1997), kombewa reduction sequence (Owen, 1938a) unipolar, bipolar and centripetal surfaces debitage (Bordes, 1961; Otte et al., 1990; Boëda et al., 1996; de Lumley and Barsky, 2004; Vallin et al., 2006), debitage of axial plan, and debitage aimed at producing triangular flakes (Locht et al., 2003; Marciani, 2018).

The integrated types comprise pieces in which the entire volume is used as a core, i.e. the whole volume of a core is involved in the realisation of products. The core in its entirety is an integral part of a comprehensive productive synergy. Moreover, considerable effort is invested in the first phases (initialisation and configuration) of core reduction. From the very beginning of the reduction, the knapper is working to realise a specific, predetermined stone product. Integrated cores are thus able to produce a recurrent series of products following a high degree of pre-planning. The types of debitage identified as integrated are: discoid, pyramidal, Levallois and laminar (Boëda, 2013).

This distinction allows us to consider technological parameters in combination with behavioural factors. The differentiation of types is based on the evidence produced by specific human actions, by which we can describe the degree of predetermination used to obtain the target object and the degree of pre-planning involved in block selection and in the management of the flaking activity (Marciani, 2018).

Regarding integrated concepts, such as Levallois, discoid and lamellar productions, there is general uniformity, whereas interpreting the vocabulary relating to other production modes was a bit more difficult. Therefore, in order not to misrepresent data, we expanded the categories by considering as additional all the productions labelled in the literature as: SSDA, Kombewa, Kombewa *sensu latu*, opportunistic, volumetric debitage (with no management of the convexities), semi-turning debitage, orthogonal debitage, unipolar, bipolar and centripetal surface debitage. We are well aware that this is a simplification of Boëda's method, due to the fact that we had to apply it on data taken from already published papers. This admittedly runs the risk of over simplifying each assemblage's actual characteristic. However, in order to develop a synthetic overview over time and space scales as large as the ones under consideration here, we needed to apply some key parameters, which proved useful for comparing sites studied by different scholars with different approaches, and to various depth.

## 3. Mousterian

The Late Mousterian has a patchy distribution across Italy. It is characterised by the dominant use of the Levallois concept, which is widely utilized in most sites in all its modalities, especially the recurrent ones. In north eastern Italy, the Levallois unipolar modality turns to centripetal in the last exploitation stages at Fumane (Peresani, 2012) and San Bernardino (Peresani, 1996). However, an alternance of the Levallois with the discoid is reported from Fumane and Rio Secco (Peresani, 2012; Peresani et al., 2014; Delpiano et al., 2018). In the northwestern part of Italy (the Ligurian - Provencal arc), the Levallois and the discoid concepts either coexist (Bombrini, Mochi, Principe – Grimaldi and Santaniello, 2014; Rossoni-Notter et al., 2017; Negrino and Riel-Salvatore, 2018; Riel-Salvatore and Negrino, 2018b), or only the discoid debitage occurs (see, e.g. layer II Arma delle Manie; Leger, 2012). Discoid debitage is also characteristic of the Late Mousterian in some Apulian sites, such as Cavallo and Bernardini (Carmignani, 2011, 2017; Romagnoli, 2012).

Sites in Central Italy are also characterised by the prevalence of the Levallois debitage. Some of them, like Grotta Breuil and Grotta dei Santi share similar geomorphological settings, and they both exploited small pebbles using recurrent unipolar Levallois debitage limited to one or two generations of target objects (Grimaldi and Spinapolice, 2010; Grimaldi and Santaniello, 2014; Moroni et al., 2018).

The sites located in southern Italy are characterised by a predominance of Levallois debitage utilising recurrent unipolar and convergent modalities which, at the end of the reduction sequence, usually changed to a centripetal or preferential modality. This pattern is mainly found at Riparo del Poggio (Caramia and Gambassini, 2006; Boscato et al., 2009), Castelcivita (Gambassini, 1997) and at Oscurusciuto (Marciani et al., 2016, 2018; Spagnolo et al., 2016, 2018; Ranaldo, 2017; Marciani, 2018).

In the Salento, at Grotta Romanelli, the Levallois sequence follows two dominant recurrent modalities: the centripetal and the unidirectional; at Grotta Mario Bernardini the recurrent centripetal Levallois predominates, like at Uluzzo C (Spinapolice, 2018a; 2018b).

The production of blades is also known from several sites: on the one hand, as a Levallois end-product mainly at Fumane (Peresani, 2011; Gennai, 2016), Riosecco (Peresani et al., 2014), Monte Netto (Delpiano et al., in press), Mochi (Grimaldi and Santaniello, 2014), Poggio (Caramia and Gambassini, 2006), Castelcivita (Gambassini, 1997), Oscurusciuto (Ranaldo et al., 2017a; Marciani, 2018), Cavallo and Bernardini (Carmignani, 2011); on the other hand, as a unipolar volumetric debitage, at Fumane (Peresani, 2011), Madonna dell'Arma (Cauche, 2007), Grotta dei Santi (study ongoing), Grotta Breuil (Grimaldi and Santaniello, 2014) Grotta Reali (Arzarello et al., 2004) and Oscurusciuto (Ranaldo et al., 2017b). The sporadic production of bladelets is attested at the sites of Fumane (Peresani and Centi di Taranto, 2013; Peresani et al., 2016), Grotta dei Santi (Moroni et al., 2018), Oscurusciuto (Marciani et al., 2016; Marciani, 2018), and Cavallo (Carmignani, 2010). Blades and bladelets were produced by utilising both Levallois and volumetric debitage, adapting and controlling the reduction sequences to suit a variety of raw blocks, such as pebbles, slabs and nodules.

The essential characteristics of the Italian Late Mousterian can be summarised as follows.

### *3.1. Raw material procurement*

The raw material procurement exploits mostly local or circum-local sources and, exceptionally exogenous sources (Spinapolice, 2012; Delpiano et al., 2018). The reduction is applied to a great variety of block types, such as pebbles, nodules, and slabs (Fig. 5) (Table 1).

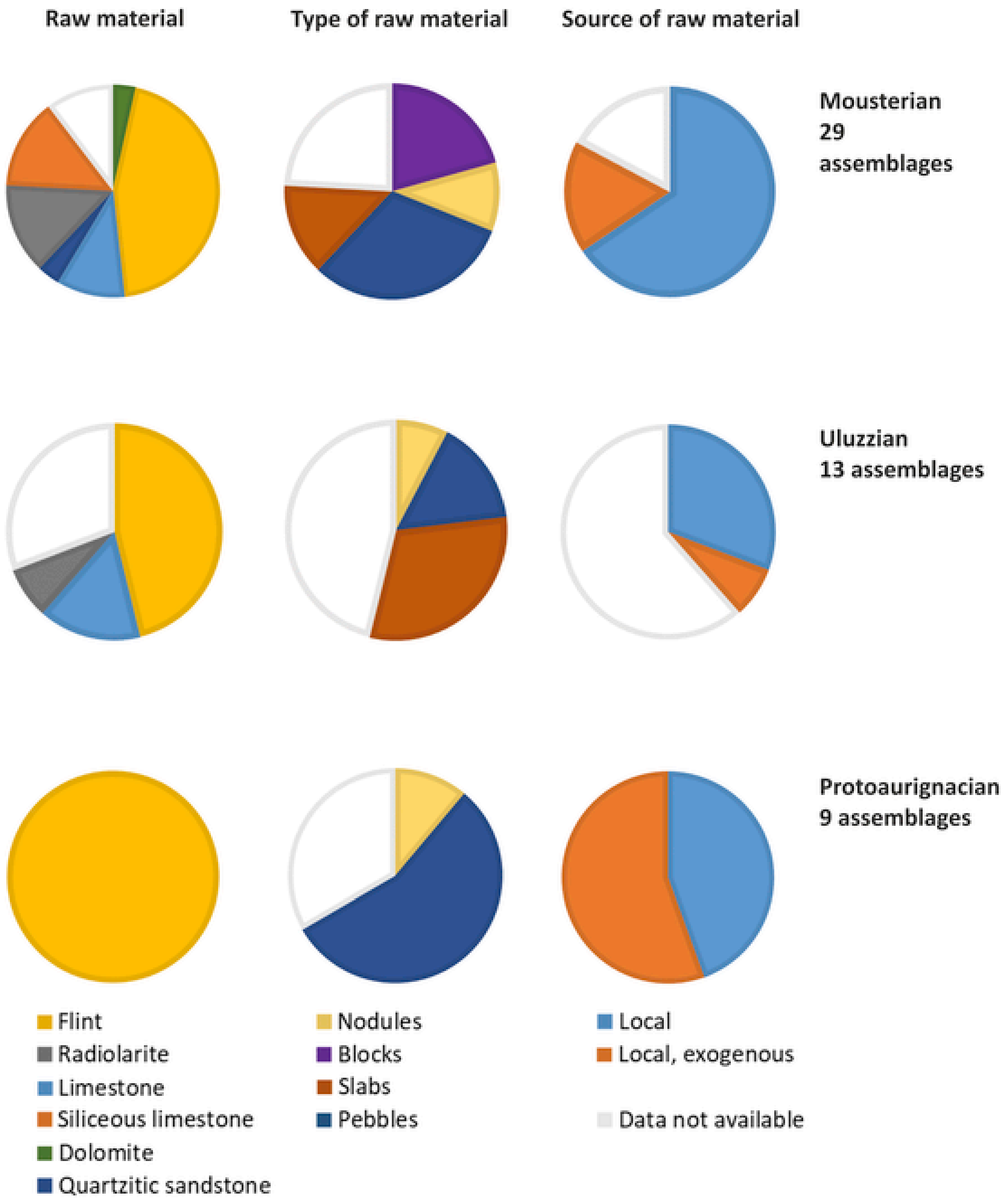

**Fig. 5.** Mousterian, Uluzzian and Protoaurignacian key sites according to the litology, type and source of raw material. Raw data and reference in Table 1.

### *3.2. Concepts of debitage*

We note the general occurrence of integrated production methods (the Levallois, and to a lesser extent, the discoid), which predominated over additional methods (unidirectional volumetric debitage, opportunistic debitage, SSDA, kombewa).

Direct percussion is carried out freehand with a hard hammerstone. In some sites, bipolar knapping on anvil also occurs, although at very low frequencies (Fig. 6) (Table 1).

### *3.3. Objective of debitage*

Production is mainly aimed at obtaining flakes, elongated supports, blades and occasionally bladelets (Figs. 6 and 7) (Table 1).

### *3.4. Retouched tools*

There is a systematic production of scrapers, mostly side-scrapers with variable evidence of reduction leading to the appearance of double converging and thinned types (Figs. 6 and 7) (Table 1).

Notably, the major technical effort occurs mostly in the production phase and not in the transformation/curation phases.

## 4. Uluzzian

The Uluzzian was initially identified by A. Palma di Cesnola (1963, 1964) at Grotta del Cavallo, Uluzzo bay (Salento, Apulia), in 1963-64. Between 1963 and 2004 (Palma di Cesnola, 2004bRiel-Salvatore, 2010), he published extensively on the new techno-complex, using then-current methods (i.e., largely the Laplace typol-ogy). Since then, there has been a certain confusion in the literature regarding the technological features of the Uluzzian, this being essentially due to two related factors: a) the use of the fundamentally typological studies of Palma di Cesnola as a point of reference, owing to the absence of an exhaustive modern revision of the Uluzzian lithic material in general; b) the attribution to this cultural entity of layers A3 and A4 of Grotta di Fumane, regardless of the clearly Mousterian component shown overall by layer A4 (Peresaniet al., 2016). This latter component is reported by Peresani et al. (2016, 2019) as a key aspect of the earliest Uluzzian (for an opposing view see Palma di Cesnola, 1993 and Moroni et al., 2018a). Therefore, despite the attribution in 2011 of the two human teeth from the lowermost layer EIII of Grotta del Cavallo to MHs (Benazzi et al., 2011), the Uluzzian has been assigned to the list of the transitional complexes *sensu* Hublin (2015 and references therein), namely those assemblages displaying a co-occurrence of Middle and Upper Palaeolithic characteristics and thus considered as expressions of last Neandertals. Moreover sharp criticism has been reported on the association between human teeth of the layer EIII at Grotta del Cavallo (uncontroversially attributed to *Homo sapiens* (Benazzi et al., 2011)) and the stratigraphic association with Uluzzian artefacts (Banks et al., 2013; Zilhão et al., 2015), critiques that have been the subject of detailed rebuttals that hopefully have definitively clarified such issue (Moroni et al., 2018a; Ronchitelli et al., 2018).

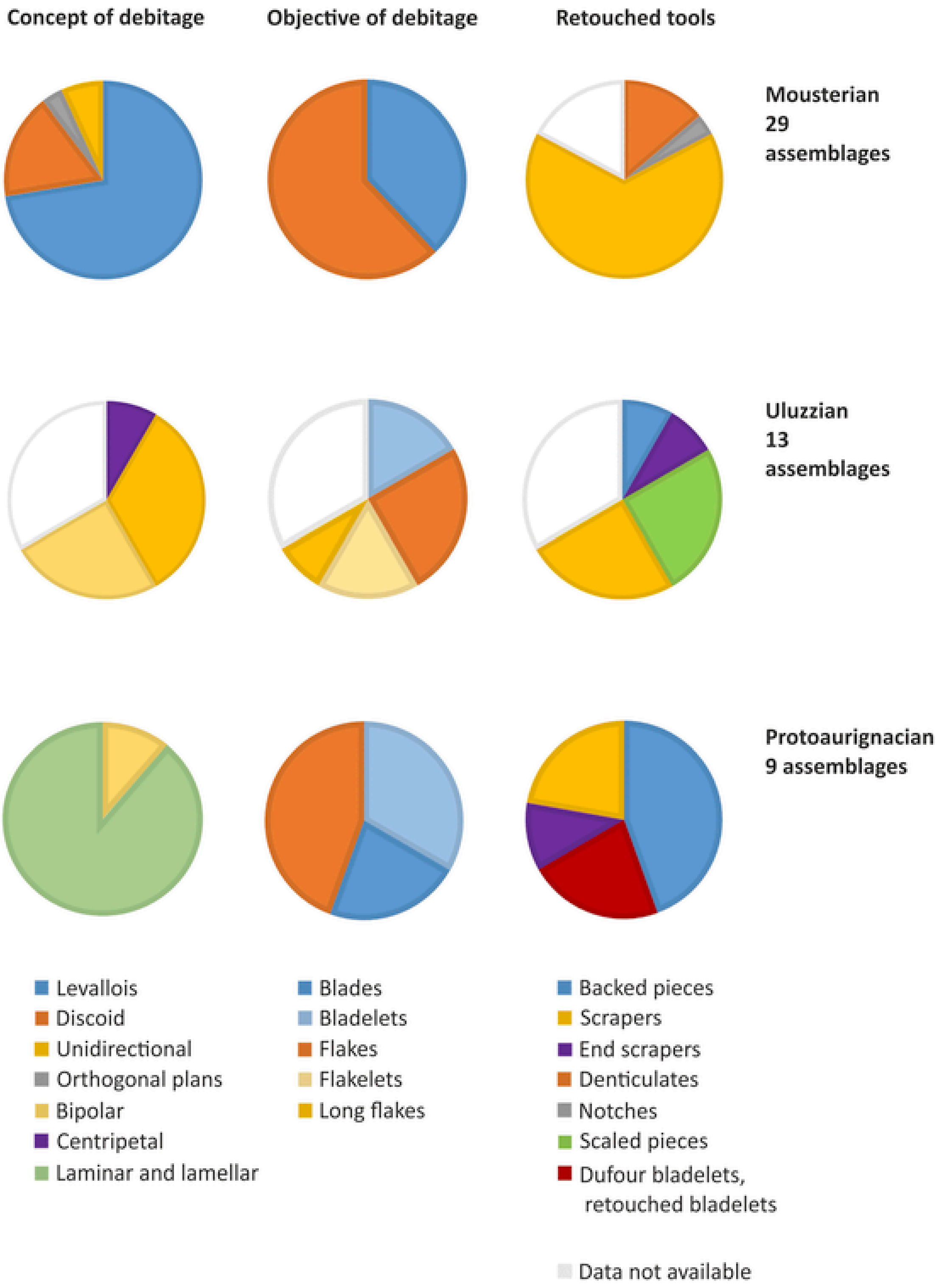

**Fig. 6.** Mousterian, Uluzzian and Protoaurignacian key sites according to the concept and objective of debitage and the most represented retouched tools. Raw data and reference in Table 1.

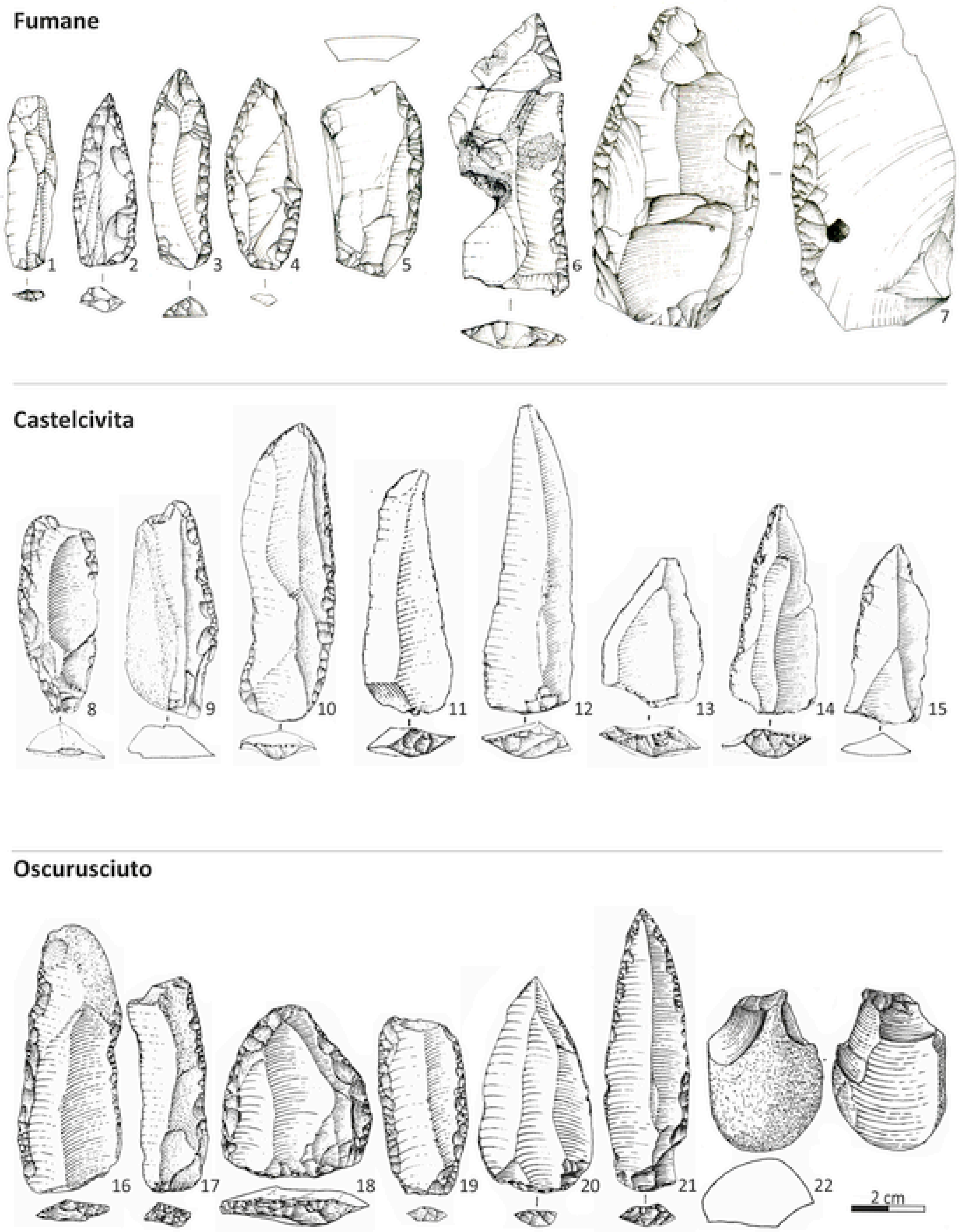

**Fig. 7.** Key Mousterian lithic artefacts. Fumane (1–7, layers A5-A6 drawings by G. Almerigogna) 1: unretouched blade; 2: retouched point; 3–7 side-scrapers. Castelcivita (8–15, layers rsi-gar-cgr; drawings by G. Fabbri; modified from Gambassini, 1997) 8–10 side-scrapers; 11–13 un-retouched Levallois points; 14, 15 retouched Levallois points. Oscurusciuto (16–22, layers 1–4 drawings by G. Fabbri and A. Moroni) 16, 19 double side-scrapers; 17 side-scraper; 18, 21 retouched Levallois points; 20 un-retouched Levallois point; 22 bladelet core.

Only very recently it has become possible to dispel a set of preconceptions (about the Uluzzian), including the occurrence of a combination of MP and UP technologies. This was the result, first of all, of a detailed revision of a large sample of the lithic assemblage from Grotta del Cavallo, layer EIII (Moroni et al., 2018a) and from Grotta di Castelcivita carried out under the aegis of the European project SUCCESS. As one of the results of this revision and of the discovery of a further Uluzzian site investigated in the North of Italy, Riparo del

Broion (Peresani et al., 2019), the nature of layer A4 of Fumane (the layer supporting the origin of the Uluzzian as rooted in local Mousterian (Peresani et al., 2016)), has been reconsidered as this layer proved to be more consistent with Mousterian characteristics. Furthermore, the chronological overlap recorded between Riparo del Broion and Fumane layer A3 points to a deviation of the latter from the more typical Uluzzian outline, which needs to be investigated in depth in the future (Peresani et al., 2016; 2019). The arrival of the Uluzzian marks a sharp break with the preceding and partially coeval Mousterian techno-complex. Data presented here are inferred both from past (Palma di Cesnola, 1993; Gambassini, 1997) and more recent sources (Ronchitelli et al., 2009, 2018; Riel-Salvatore, 2009, 2010; Boscato et al., 2011; Boscato and Crezzini, 2012; De Stefani et al., 2012; Douka et al., 2012, 2014; Wood et al., 2012; Moroni et al., 2013, 2016; 2018a; Villa et al., 2018; Peresani et al., 2019) as well as from ongoing studies.

The assemblage from Levels 1g, 1f at Riparo del Broion shows a high fragmentation rate caused by the use of bipolar knapping technique, and a lamino-lamellar production (Peresani et al., 2019).

La Fabbrica is characterised by cores with a flat striking platform opened by a single scar or by previous scars orthogonal to the debitage surface which present unidirectional parallel removals. The exploitation could involve only one or two adjacent debitage surfaces. Sometimes orthogonal removals on one or two debitage surfaces have been recorded together with an abundant bipolar component (Villa et al., 2018).

At Colle Rotondo unidirectional, bidirectional or multidirectional cores with parallel removals on one, or more surfaces of debitage, are very common. The striking platform can be cortical or opened by one or several removals (Villa et al., 2018). Again the bipolar technique is reported as being the dominant reduction strategy.

Uluzzo C is characterised by the production of bladelets and flakelets produced by both volumetric debitage with a direct percussion technique and another debitage resulting from the use of the bipolar technique . This same co-occurrence of these two components has also been noted in the Uluzzian assemblages at Castelcivita.

These sites which till now have been studied by a technological approach show some internal variability (possibly due to the different chronological phases that they represent or to different local adaptation) in the mode of production but at the same time several common features can be underlined. We note that the Levallois and discoid debitage which characterised the Mousterian are missing in the Uluzzian (except at Fumane (Peresani et al., 2016, 2019). It is clear that the presence of a bipolar production and a volumetric debitage produced by direct percussion define the techno-complex. However, it is necessary to better define these two components, to see whether they are part of the same sequence or rather represent two distinct reduction sequences.

The main characteristics of the Uluzzian lithic industry can be summarised as follows:

- extensive use of the bipolar technique with systematic production of flakes and blades of very small dimensions;
- a volumetric debitage with a direct percussion technique; management of the striking platform and of the lateral and distal convexities is mainly present in the last phases;
- absence of Levallois and discoid debitage;
- low technical effort in the production phase;
- occurrence of new tools: the lunates;
- systematic production of short end-scrapers on slabs and flakes.

### *4.1. Raw material procurement*

Apart from the north Italian contexts where changes in raw material procurement were influenced by the availability of knappable stone, the procurement of raw material remains generally confined, like for the Mousterian, to local and circum-local sources (Dini and Tozzi, 2012; Villa et al., 2018; Moroni et al., 2018a) although an increase in possibly exogenous flint has been noted from the archaic to the final phases in Salento (Ranaldo et al., 2017a) (Fig. 5) (Table 1).

### *4.2. Concepts of debitage*

The integrated production concepts (i.e. Levallois, discoid) typical of the Mousterian are lacking. On the other hand, additional debitages prevail (i.e. unipolar volumetric debitage). Among percussion techniques, the bipolar knapping on anvil is the most used, combined with the unipolar direct freehand percussion technique. From the archaic to the late/final Uluzzian, the use of bipolar technique decreases and laminar products are also obtained from ad hoc partially prepared cores (Fig. 6) (Table 1).

Bipolar products have peculiar documented characteristics (Barham, 1987; Knight, 1991; Guyodo and Marchand, 2005; Bietti et al., 2010; Bradbury, 2010; Soriano et al., 2010):

- sheared bulbs of percussion;
- shattered butts, or they are reduced to a point or a line;
- the longitudinal profile of the ventral face is generally rectilinear;
- the ventral and dorsal faces often very similar;
- the ventral face sometimes characterised by very pronounced ripple marks.

### *4.3. Objective of debitage*

The typical products resulting from bipolar reduction are thin and straight small flakes and flakelets, sometimes smaller than 1 cm, and small blades/bladelets. Other distinctive products are thick blades/small blades with quadrilateral cross-sections. Many of the bipolar products, especially those of very small dimensions, are supposed to have been hafted "as is" without any modifications (see for more details Riel-Salvatore, 2009; Moroni et al., 2018a). As the Uluzzian evolves over time we note an increase in the production of blades (mainly small blades and bladelets) which generally become more standardised (Figs. 6 and 8) (Table 1).

### *4.4. Retouched tools*

Formal tool sare mostly backed pieces (mainly lunates), short end-scrapers and side-scrapers as well as some denticulates. At Cavallo, marginally backed small blades with irregular profiles, produced through the bipolar technique (unlike classic Dufour bladelets), have sporadically been found in layer EIII. Furthermore, a few marginally backed bladelets are present in layer D (study ongoing).

The occurrence at Grotta del Cavallo and Castelcivita of some flattened sandstone pebbles used as anvils must also be noted.

Due to the characteristics of raw material (silicified limestone slabs), a peculiar tool production, distinctive of the Salento region (mainly during the archaic phase) consists of directly employing thinner (15-5 mm) naturally fragmented slabs (lastrine) as blanks for retouched tools without any previous debitage modification (Figs. 6 and 8) (Table 1).

## 5. Protoaurignacian

The Protoaurignacian has been considered one of the cultural manifestations of the initial MH migration into Europe (Bailey and Hublin, 2005; Mellars, 2006; Nigst et al., 2014; Benazzi et al., 2015). It appears over a vast geographic region including Italy (e.g. Fumane (Bartolomei et al., 1992; Broglio et al., 2005; Bertola et al., 2013; Falcucciet al., 2017), Mochi (Kuhn and Stiner, 1998; Bietti and Negrino, 2008; Grimaldiet al., 2014), Bombrini (Bietti and Negrino, 2008; Bertola et al., 2013; Negrino et al., 2017; Holt et al., 2018), La Fabbrica (Dini et al., 2012), Paglicci (Palma di Cesnola, 2006) Castelcivita (Gambassini, 1997) and Serino (Accorsi et al., 1979). The Protoaurignacian is characterised by technological innovations in the lithic production and by the abundance of bone tools (awls and needles), ochre and personal ornaments (including numerous perforated shells) (for an updated review on ornaments see Arrighi et al., in this Special Issue).

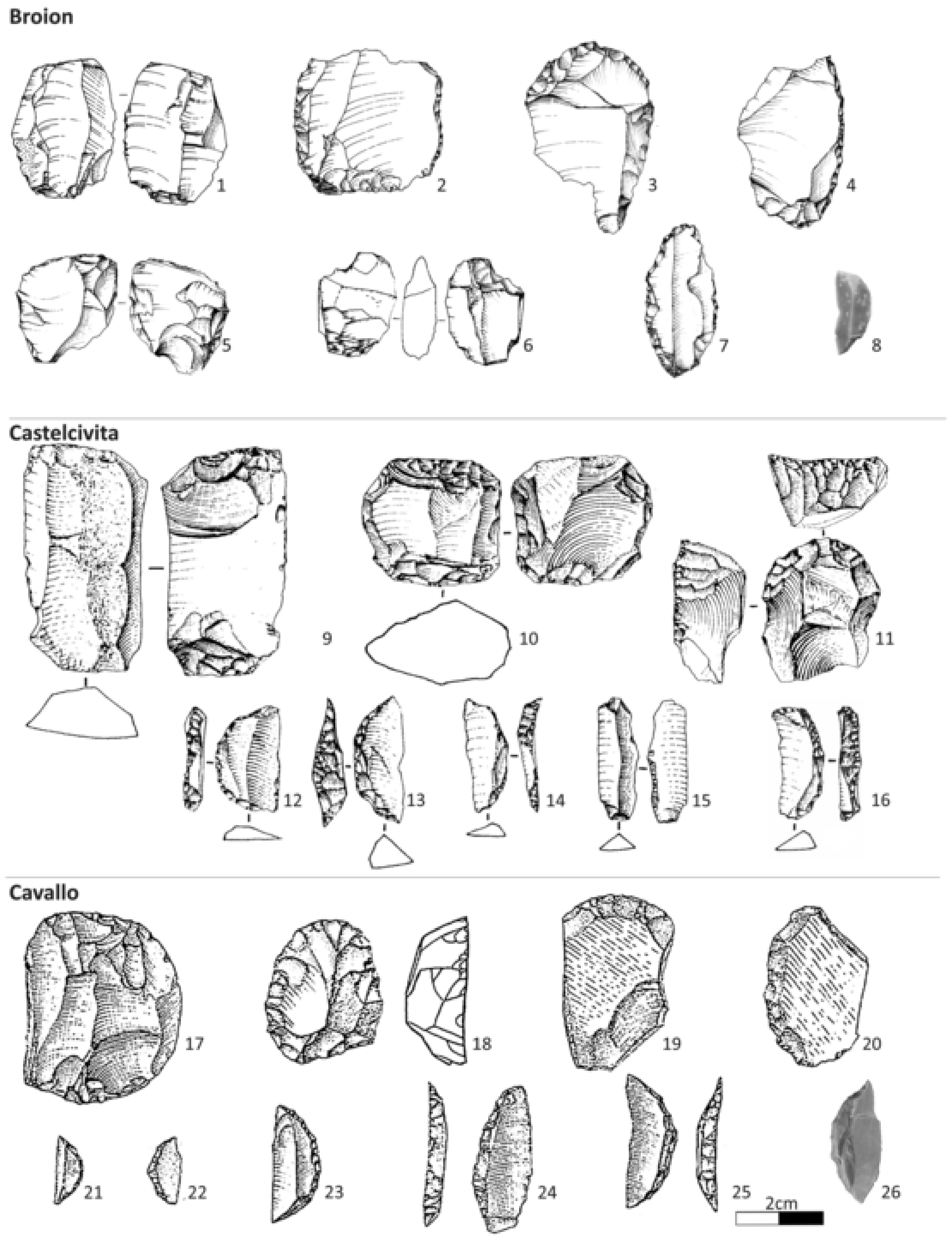

**Fig. 8.** Key Uluzzian lithic artefacts. Broion (1–8, layers 1f, 1g; drawings by G. Almerigogna, photo by D. Delpiano) 1, 2, 5, 6: splintered pieces\bipolar cores; 3: end-scraper, successively splintered; 4, 7: backed pieces; 8: lunate. Castelcivita (9–16 layers rsa'', rpi, pie drawings by G. Fabbri; modified from Gambassini, 1997) 9, 10: splintered pieces\bipolar cores; 11: end-scarper\bladelet core; 12–14,16: lunates; 15: bladelet. Cavallo (17–26 layer D, drawings by G. Fabbri, photo by S. Ricci; modified from Ranaldo et al., 2017a) 17: splintered piece\ bipolar cores; 18: end-scraper; 19, 20: end-scrapers on slab; 21–26 lunates.

The lithic typology of the Protoaurignacian was first defined by Laplace (1966) and it is characterised by the presence of Dufour bladelets (that is to say straight, elongated bladelets subsequently modified by direct inverse or alternate retouch), variably associated with carinated tools. In contrast, the Early Aurignacian is mainly characterized by the abundance of carinated tools and Aurignacian blades (de Sonneville-Bordes1960; Peyrony, 1934; de Sonneville-Bordes1960; Laplace, 1966).

Based on technological studies the Protoaurignacian has been argued to be characterised by a unique continuous production sequence aimed at producing both blades and bladelets, with bladelets occurring at the end of the reduction (Bon, 2002; Teyssandier, 2006, 2007, 2008). This is in contrast with the Early Aurignacian, in which blades and bladelets are produced by two distinct reduction sequences, with blades knapped from unidirectional prismatic cores and bladelets and microblades primarily obtained by the exploitation of carinated 'endscraper' cores (Bon, 2002; Teyssandier, 2007; Teyssandier et al., 2010). It is also assumed that these two production strategies correspond to distinct *savoirs-faire* and responded to different consumption requirements (i.e. blade for domestic tools and bladelets as armatures) (Tartaret al., 2006).

Recent studies have, however, questioned the technological and/or typological basis for separating the Protoaurignacian and Early Aurignacian into two technological traditions, based only on typological or/and technological studies. Indeed, Tafelmayer (2017) and Bataille et al. (2018) have argued that without refitting or distinguishing Raw Material Units, it is simply not possible to define whether blades and bladelets are the result of one or two reduction sequences. Moreover, they claim that separating the Protoaurignacian and Early Aurignacian in this way oversimplifies the archaeological reality which is much complicated and requires a multi-proxy model that transcends mere techno-typological systematics to reach conclusive interpretations (Bataille et al., 2018).

From a chronological standpoint, Banks et al. (2013) have proposed a model of diachronic continuity and internal evolution from the Protoaurignacian to the Early Aurignacian (see also Le Brun-Ricalens et al., 2009; Teyssandier, 2007). However, recent data have shown that this pattern does not hold at least in Central Europe as the Protoaurignacian and the Early Aurignacian overlap (Szmidt et al., 2010; Douka et al., 2012; Higham et al., 2012; Nigst and Haesaerts, 2012; Nigst et al., 2014). This model (Banks et al., 2013) is incorrect also for Italy as the Proto-Aurigancian at the Balzi Rossi and northern Italy lasts past HE4 (Riel-Salvatore and Negrino, 2018a and b). This overlap could be interpreted in two ways: 1. The Protoaurignacian and Early Aurignacian could represent different developmental trajectories, respectively the southern and northern dispersal routes of MH within Europe (Mellars, 2006); or 2. They could be different manifestations of the same general adaptive package related to the exploitation of disctinct niches requiring different food-acquisition technologies (Nigst et al., 2014). In Italy the earliest Protoaurignacian assemblages come from the Ligurian sites of Mochi (Bertola et al., 2013) and Bombrini (Riel-Salvatore, 2007; Riel-Salvatore and Negrino, 2018a and b) and from Fumane (Falcucci et al., 2017, 2018; Falcucci and Peresani, 2018) in northeastern Italy. Only after 40 ka cal BP is the Protoaurignacian documented in southern Italy, for instance at Paglicci (Palma di Cesnola, 2004b) and at Grotta della Cala (Benini et al., 1997).

The basic characteristics of the Italian Protoaurignacian can be summarised as follows:

- bladelet dominated industries with major technical effort involved in the production phase compared to the Uluzzian;
- bladelets have straight profiles and are mainly transformed in marginally backed implements. Retouch, direct or inverse, can be located on one or both edges.
- standardisation of products.

### 5.1. Raw material procurement

Compared to the Late Mousterian and the Uluzzian, there is a marked increase in the use of exogenous raw material which could also come from sources located several hundred km from the sites. This is true especially for Liguria (Riel-Salvatore and Negrino, 2009; Holt et al., 2018) but not for regions in which high-quality raw materials were available, as for instance near Fumane or Paglicci (Fig. 5) (Table 1).

### 5.2. Concepts of debitage

The incidence of bipolar knapping on anvil drops relative to the Uluzzian, and, except for La Fabbrica, where the production of flakes is dominant (Dini and Tozzi, 2012), there is a clear dominance of systems aimed at obtaining a series of standardised laminar products which involve two main reduction sequences.

Core reduction is based on two distinct operational concepts:

1. A linear and consecutive knapping progression aimed at obtaining blades and, to a lesser extent, bladelets with sub-parallel edges;
2. An alternating knapping progression exclusively used to produce slender bladelets with a convergent shape (Paglicci and Fumane) (Borgia et al., 2011; Falcucci et al., 2017, 2018).

At Mochi and Bombrini, the occurrence of a consecutive knapping progression cannot be clearly verified as blade production is completely lacking in situ, probably due to the absence of suitable raw material (like at La Fabbrica); indeed, blades are made from non-local lithotypes. Systematic crest modelling has been observed at Mochi (Grimaldi and Santaniello, 2014), Bombrini (Riel-Salvatore, 2007) and Fumane (Falcucci and Peresani, 2018). At Bombrini, beside the production of bladelets there is also a debitage geared at producing flakes including some elongated blade-like blanks. The flake production is an important part of the assemblage and seems to be a secondary product of blade production as core reduction advanced (Riel-Salvatore and Negrino, 2018a).

Both hard and soft hammers have been documented (Borgia et al., 2011; Caricola et al., 2018) (Fig. 6) (Table 1).

### 5.3. Objective of debitage

The target products are usually blades, small blades and bladelets (Table 1). Curved profiles of different grades clearly dominate the blades, whereas straight profiles are more common among bladelets. Twisted items are rare. Their edges are sub-parallel or convergent. Dimensions cannot be easily calculated because of the high fragmentation, but products probably measured no more than a few cms (Figs. 6 and 9) (Table 1).

### 5.4. Retouched tools

The retouch of bladelets, direct, inverse or alternate, is always semi-abrupt and continuous. Other typical tools are the marginally backed points on bladelets retouched on both edges. These are common in the north at Fumane (Falcucci et al., 2017), Bombrini (Negrino et al., 2017) and, in the form of micro-points, in the uppermost layers of Castelcivita, in Campania. Blades were selected to manufacture end-scrapers, burins and laterally-retouched tools. Among end-scrapers, carinated implements of varying thickness predominate (Figs. 6 and 9) (Table 1).

## 6. Discussion

The last phase of the Mousterian is documented at many sites throughout Italy, and it is mostly characterised by integrated debitage concepts, i.e. the Levallois and the discoid. The largely prevailing use of integrated methods indicates that the major effort in the Mousterian industries was focused on the production phase, i.e., that there was an investment in the initialisation of the block and in maintaining determinate convexities and characteristics of the block until the end of the production. The transformation phase (retouching) appears to be less important (Table 1 and references therein).

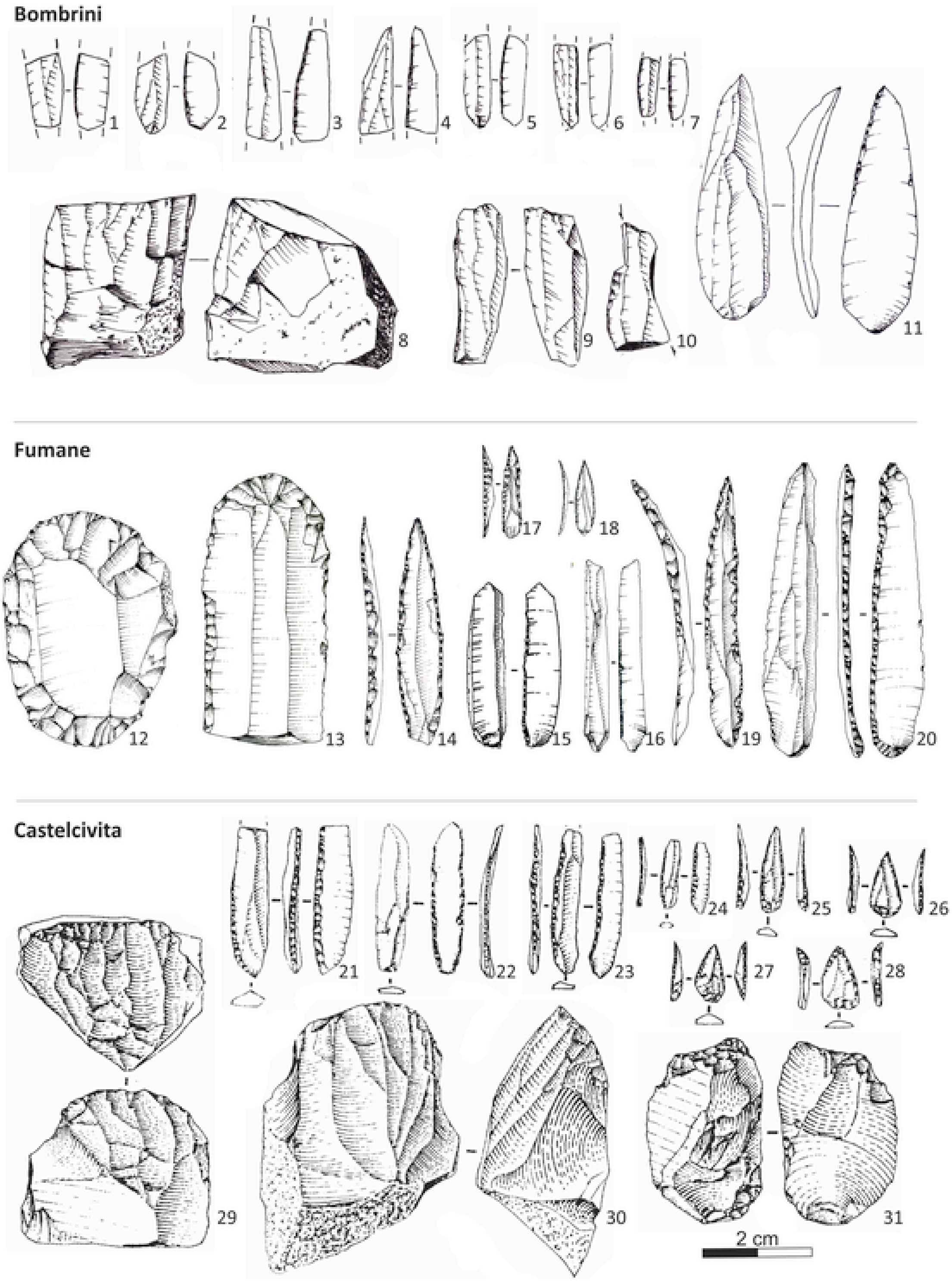

**Fig. 9.** Key Protoaurignacian lithic artefacts. Bombrini (1–11 layer A1 drawings by F. Negrino; modified from Negrino and Riel-Salvatore, 2018) 1–7: Dufour bladelets; 8, 9 bladelet core; 10 burin; 11 point with marginal retouch. Fumane (12–20 layer A1-A2 drawings by A. Falcucci) 12,13: end-scrapers; 14, 19, 20: retouched points; 15: Dufour bladelet; 16: un-retouched bladelet; 17, 18: retouched micro-points. Castelcivita (21, 31 layer ars, cgr, drawings by G. Fabbri; modified from Gambassini, 1997). 21–24: Dufour bladelets; 25–28: retouched micro points; 29, 30: end-scarper\bladelet core; 31: splintered piece\bipolar cores.

Uluzzian lithic technology can be seen as a clear-cut rupture with this previous reality in that, unlike the Mousterian, it is characterised by the use of non-integrated production systems and consequently by a low focus on the production process. It should be emphasised that this does not necessarily mean a general lack of technological complexity and that technical innovations introduced in the Uluzzian likely went far beyond lithic production. It is well known that stone tools are often only a part of the entire technological system comprising the design of implements and weapons which includes the necessary know-how related to ballistics, hafting, fletching etc … In the case of the Uluzzian, this is apparent as shown by the occurrences of lunates displaying clear traces of impact fractures, suggesting their use as armatures in throw-ing weapons (Sano et al.Sano et al., 2019). A similar use can be supposed for flakelets and bladelets produced by bipolar technique, as hinted at by some ethnographic, archaeological and experimental instances (White and White, 1968; Chauchat et al., 1985; Shott, 1989; Crovetto et al., 1994; Le Brun-Ricalens, 2006; Riel-Salvatore, 2009; de la Peña et al., 2018; Moroni et al., 2018a), and underpinned by the very preliminary results from the use-wear studies carried out on a few elements from Cavallo, Castelcivita and Uluzzo C.

Lithic technology is a proxy for human behaviour as well. Bipolar technique has been commonly recognised as an "expedient" production system used to save time and energy during possible "crisis" conditions (Callahan, 1987; Shott, 1989; Jeske, 1992; Hiscock, 1996; Diez-Martín et al., 2011; Mackay and Marwick, 2011; Eren et al., 2013; Morgan et al., 2015). In the archaic Uluzzian, bipolar knapping is associated with the extensive use of a very singular technique in making tools –mostly end-scrapers and side-scrapers – by directly retouching thin slabs without any previous debitage. For the Salento region, this pronounced reliance on "low-cost techniques", combined with the extensive exploitation of local or nearly local raw material, besides being symptomatic of a very low technical effort during the production phase, has been interpreted (Moroni et al., 2018a) as a possible sign of reduced mobility by the Uluzzian groups, due to insufficient territory expertise/control or to low demographic density/instability or both.

According to radiometric and geochronological determinations, the Uluzzian makes its appearance, develops and dies in the space of about 5000 years. The initial roots of the Uluzzian are still a matter of investigation and of fervent debate. Apart from the specific case of Fumane (Peresani et al., 2016, but cf. Moroniet al., 2018a), current data indicate an absence of evidence of roots for the Uluzzian in the local Late Mousterian, considerable technological innovations of the Uluzzian as well as its association with MH remains. All of these elements are congruent with a non-local origin of this techno-complex. The late and the final phases of the Uluzzian assemblages inSouthern Italy display an increasing occurrence of Aurignacian items and a decline of the most typical features of the Uluzzian. Based on the evidence provided by layer D of Cavallo and layer B of Serra Cicora, Palma di Cesnola assumed that this phenomenon was perhaps symptomatic of a gradual "cultural hybridisation" between the two cultures (Palma di Cesnola, 1993, p. 150), finally resulting in the assimilation of the last Uluzzian groups by Aurignacian groups when they reached southern Italy after 40 ka cal BP. Very preliminary results of the techno-typological and taphonomic revision carried out on layer D of Cavallo allow us, for the moment, to rule out even minor post-depositional disturbances as a factor for the latest Uluzzian's distinctive techno-typological features. If Palma di Cesnola's hypothesis is confirmed by ongoing studies, this would provide an explanation for the sudden disappearance of the Uluzzian techno-complex from the Italo-Balkan region around 40 ka ago.

The Protoaurignacian is found in large parts of Europe, from the Balkans to the Spanish Mediterranean coast and into Cantabria. According to the available radiometric chronology (Table 1 SM), the Protoaurignacian makes its appearance in Italy about 42-41 ka cal BP at Riparo Mochi (Liguria; Douka et al., 2012) and at the penecontemporaneous site of Fumane (layer A2) (about 41 ka cal BP; Higham et al., 2009 ). Data available for northern Italy do not allow any inference about a possible dispersion route of the Protoaurignacian from east to west or vice versa. On the contrary, a rapid north-south diffusion of this techno-complex across the Peninsula is suggested by the younger age ranges (about 40 ka cal BP) of the southern assemblages.

There is no technological continuity between the Protoaurignacian and the Mousterian, while some connection can be assumed, as said above, between the Protoaurignacian and the Uluzzian in the Salento region, where the late/final phases of the Uluzzian yielded a certain amount of Aurignacian-like pieces which possibly attest to contacts occurring between the two populations.

The main novelty of the Protoaurignacian is represented by the production of standardised bladelets used as blanks for the marginally backed Dufour bladelets, for the backed points of Fumane and for the backed micro points of Castelcivita. The function of these particular points has not been investigated in any depth, and the hypothesis of their use as armatures in weapons (Broglio et al., 1998) must be validated by future targeted use-wear studies. Large blades were predominantly used to obtain domestic tools such as side-scrapers, end-scrapers and burins, whose carinated variants can likely be considered as cores on flakes aimed at producing bladelets.

In a few sites of northern and central Italy (Riparo Mochi-Layers F ed E, Grotta dei Fanciulli-Layers K ed I, Fumane Layer D, Fossellone Layer 21), the period examined so far is followed by new lithic assemblages sharing several Aurignacian-type features: carenated and nosed end-scrapers, *busqué* burins, twisted retouched bladelets (of the Roc-de-Combe type) and deeply and invasively retouched large blades. These are found in association with split-based bone points.

In southern Italy, at Grotta Paglicci (where the CI is absent), the Protoaurignacian seems to expand chronologically (layer A1-0; Palma di Cesnola, 2004b) with an original aspect characterised by the occurrence of asymmetric twisted *déjetées* bladelets. Both in this Adriatic site and on the opposite Tyrrhenian side, at Grotta della Cala (where the CI is also absent), the classic Aurignacian is lacking and, after a hiatus in sedimentation, the Protoaurignacian is overlain by the ancient Gravettian dated to about 30 ka cal BP. Technological, chronological and environmental studies would be needed to disentangle the question concerning the lithic assemblage retrieved in stratigraphy only at the cave of Serra Cicora (horizons A, B and C of layer B), in the Salento (Spennato, 1981), located 1.6 km from Grotta del Cavallo as the crow flies (Fig. 1 n. 35). This assemblage has been compared to some surface collections from Calabria and Tuscany and referred to by Palma di Cesnola as a phylum of the Aurignacian that he called "Uluzzo-Aurignaziano" (1993), suggesting a possible "hybridisation" between the two technocomplexes.

Focusing on lithic technology, what really differentiates the three techno-complexes is the concept of debitage, which encompasses the ways of reducing the raw material and producing desired end-products (Table 1) (Fig. 10). During the Late Mousterian, the Levallois in its unipolar and centripetal forms is the dominant concept, even though also the discoid production plays a part (Fig. 10). According to Boëda (2013) both the Levallois and the discoid are integrated concepts of debitage, which means that there is a strong focus on the choice of the initial raw material block and an investment in the management of the convexities to produce specific kinds of blanks, such as flakes, elongated flakes and blades.

The Uluzzian is characterised by the application of additional concepts, that is to say debitage where the striking platform is a natural or cortical plan, or where it is created by a single or few removals, and where one or more side of the core are used independently as debitage planes. Knapping strategies are dominated by the bipolar technique and mainly geared towards the production of small blades/bladelets and small flakes/flakelets (Table 1) (Fig. 10). There thus appears to be a clear change in the types of desired end-products, in terms of techno-dimensional categories. Whereas the Uluzzian and the Protoaurignacian lithic sets are characterised by two distinct size components (larger-size-tools and smaller-size-tools) which usually correspond to different functional activities (Moroni et al., 2018a), this dichotomy is absent in the Late Mousterian.

The dominance of additional debitages in the Uluzzian results in a low effort in the management of convexities and striking platforms. The block is not exploited in its entirety (in contrast to the integrated concept), which represents a crucial change in the use of volumes of raw material.

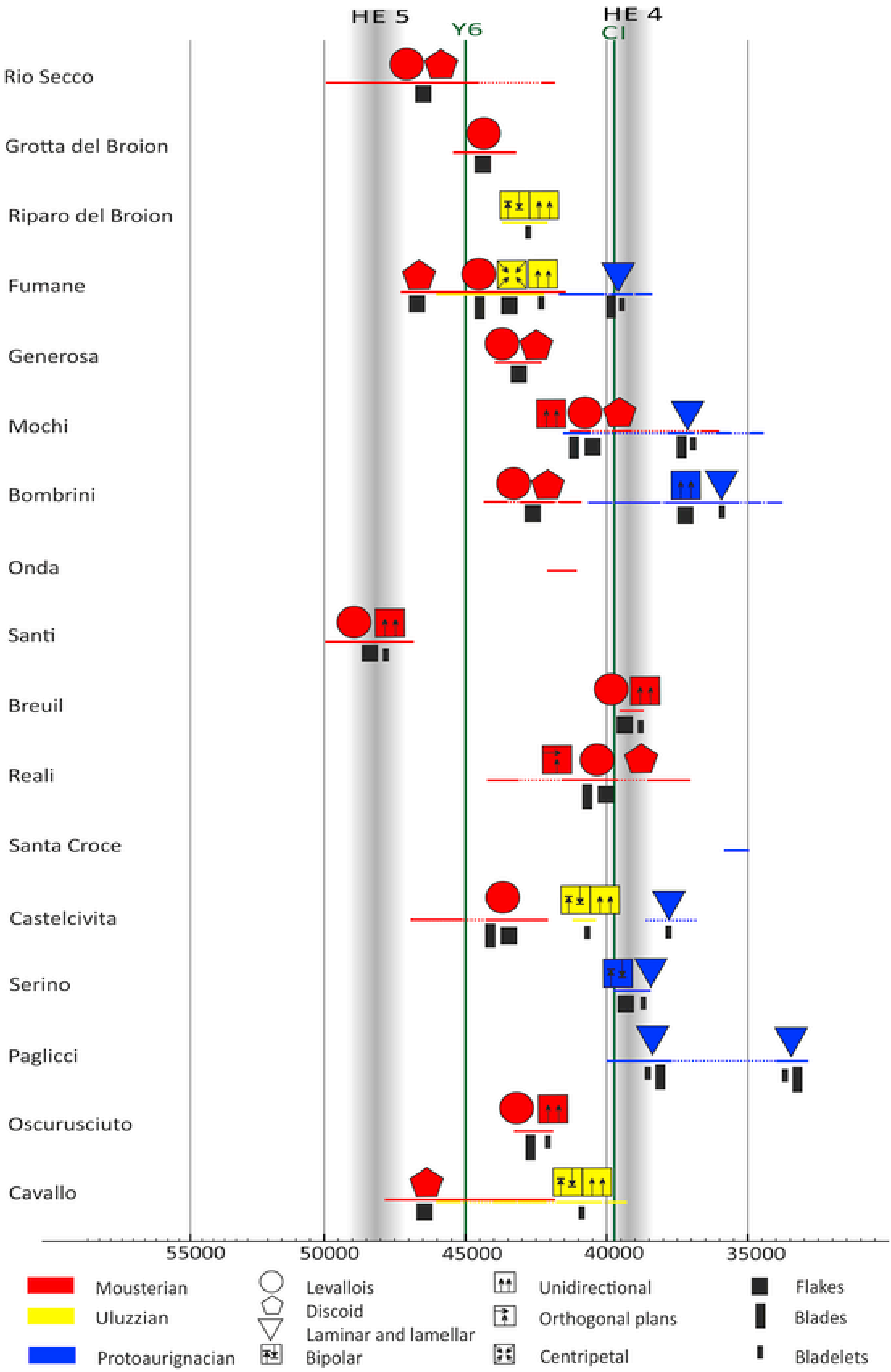

**Fig. 10.** Overview of Mousterian, Uluzzian and Protoaurignacian lithic behaviour according with the most utilized concept of debitage and objective of flaking. 14C raw dates and bibliographic references in Table 1 SM, bibliographic references of concept and objective of debitage in Table 1.

We also note the presence of diverse techniques of debitage: both direct and bipolar percussion are present, executed using hard hammers.

The bipolar technique is already present during the Late Mousterian, where it plays a consistently marginal role (e.g., in Fumane A4, Cavallo FII-I) and is sometimes functionally exclusive, for instance when it is used solely to'open' small pebbles at Grotta Breuil. In the Uluzzian, in contrast, the bipolar knapping is a systematic technical choice that is implemented extensively and applied indifferently to all types of available units of raw material (such as block, nodule, slab and pebbles).

In the Protoaurignacian, integrated concepts predominate, namely the volumetric blades and bladelet debitages characterised by a great effort in managing the distal and lateral convexities, as well as in preparing striking platforms (Table 1) (Fig. 10). We also note a strong control of the angles, which allows the knapper to continuously produce blanks until the utility of a given core is exhausted. Protoaurignacian knappers used both direct and bipolar percussion performed with a variety of hammers (i.e. soft, hard, organic.), depending on the stage of the debitage.

Although the Uluzzian and the Protoaurignacian makers employed radically different concepts of debitage (respectively, additional [i.e. volumetric and orthogonal debitage] and integrated concepts [i.e laminar, and lamellar debitage via unidirectional and prismatic cores]), it should be noted that these two groups mainly pursued the production of similar end-products (i.e. blades, small blades and bladelets), perhaps suggesting comparable needs and behaviours.

Laminar and occasionally lamellar production in some assemblages at the end of the Middle Paleolithic is a trend registered in Italy (e.g. Peresani, 2012; Gennai, 2016; Carmignani, 2017, 2018; Marciani, 2018) (Fig. 10) and elsewhere in Europe (e.g. Révillion and Tuffreau, 1994; Bar-Yosef and Kuhn, 1999; Maíllo Fernándezet al., 2004; Slimak and Lucas, 2005; Pastoors, 2009; Pastoors and Tafelmaier, 2010; Zwyns, 2012a Zwyns, 2012b). The way of producing blades and bladelets and the role of this kind of production may have played in the behavioural dynamics of the transition to the Upper Palaeolithic is becoming a hot topic in scientific debate.

It has been observed that there is increase in the number of bladelets produced, that is to say they are present in the Mousterian, abundant in the Uluzzian and finally reaching a dominant role in the Protoaurignacian. We can also note that the production systems used to obtain these blanks change, and become more standardised and efficient. In the Mousterian, however, bladelets (and blades as well) appear to represent just two blank types among several others (Pastoors, 2009). The key question is why do we at a certain point observe the production switch from producing flakes, blade and bladelets to a targeted, standardisedand almost “industrial” production of bladelets. Assuming that the production of a tool is driven by a specific necessity, this would suggest that, in the Upper Paleolithic, new necessities arose that, required the production of a great number of standardised bladelets, which could be used in composite tools (Hays and Lucas, 2001; Broglio et al., 2005; O'Farrell, 2005; Pelegrin and O'Farrell, M., 2005; Borgia and Ranaldo, 2009; Borgia et al., 2011).

The Late Mousterian and Uluzzian seem to have had a preference for local and circum-local material whereas the Protoaurignacian displays a greater dependence on the procurement of exogenous raw material acquired, at least occasionally as in Liguria, over very wide territories. This distinction is one of the reasons why the Uluzzian for a long time was attributed to Neandertals (Bietti and Negrino, 2007). However, more regional studies are needed to fully understand the role played by the availability of raw material and its exploitation by different groups.

## 7. Conclusion and new directions

Based on lithic data synthesised from various studies, we were able to confirm and detail the major differences between the Late Mousterian, Uluzzian and Protoaurignacian techno-complexes. These differences include not only typological distinctions, but also important differences in technical and technological aspects related to the way of conceiving and making tools. What is needed at this point to move the debate forward are in-depth investigations into the root causes behind these different behavioural patterns, in terms of elements like, for instance, mental templates, mobility patterns, food procurement strategies, environmental constraints, ethnic identity, demographic density, site function and technological innovation. All of these data need to further be declined against the backdrop of a high-resolution absolute chronology.

Summarising, different paths for futiure researches emerge from our review.

1. From a lithic production perspective, we have observed that the Late Mousterian principally depended on integrated reduction systems (i.e. Levallois and discoid), whereas Uluzzian lithic production was mainly additional, with less attention paid to the management phase and more emphasis put on the ‘operationalisation’ of the tools. The Protoaurignacian differs from the Uluzzian for the systematic use of integrated volumetric reduction systems (i.e. laminar). These differences in the production systems along with the different role played by retouch are worth to exploring in greater depth.
2. Considering the objectives of debitage, it is necessary to understand why there is a need to increase and standardise the production of blades and bladelets at the beginning of the Upper Paleolithic and/or what is the trigger of this technological innovation.
3. From a techno-functional point of view, the Uluzzian bladelets and flakelets should be investigated as possible inserts of composite tools.

As far as the Uluzzian is concerned, the Salento coastal belt remain a privileged region for investigating its internal dynamics in the context of the characteristics of human groups during the transition, thanks to the concentration of several sites in a restricted area. It should be underlined that identifying the timing and the modality of the disappearance or relocation of Neandertals from these sites would be of pivotal interest, since all the archaeological sequences involved display evidence of a prior Mousterian occupation. A more accurate geochronological assessment of Late Mousterian assemblages in these regions would shed new light on this issue, while also potentially providing information on population density and mobility immediately before the appearance of the Uluzzian. Even though in Italy there appears to have been a 2000–5000 year period during which the Mousterian and the Uluzzian coexisted and a 1000–2000 year overlap between the Uluzzian and the Protoaurignacian, the systematic lack of interstratification between these techno complexes probably indicates the lack of actual co-habitation among these different groups in given regions. This means that the makers of the Uluzzian and the Protoaurigancian likely occupied territories which were already devoid of Late Mousterian groups.

As a matter of fact, technological lithic studies on the Mousterian in general are in a more advanced state with respect to those of the Uluzzian and the Protoaurignacian. Therefore further studies will be crucial in filling gaps in our knowledge on lithic technological organization, and more generally on the behavioural dynamics pertinent to the replacement of Neandertals by MHs and the mutual adaptation between the two species.

Among the main aims of our project in the field of lithic studies are the definition of the technological (and ethnic) identity of the Uluzzian tool makers and their role in the peopling of Italy, as well as the investigation of the possible relationship between this population and its "next-door neighbours", namely the Mousterians (likely Neandertals) and later possibly the Protoaurignacian makers.

## Author contributions

Conceptualization: Giulia Marciani, Adriana Moroni, Annamaria Ronchitelli. Original draft: Giulia Marciani, Adriana Moroni, Annamaria Ronchitelli. Review & editing: Giulia Marciani, Adriana Moroni, Annamaria Ronchitelli, Marco Peresani, Enza Elena Spinapolice, Fabio Negrino, Julien Riel-Salvatore, Stefano Benazzi, Simona Arrighi, Federica Badino, Eugenio Bortolini, Paolo Boscato, Francesco Boschin, Jacopo Crezzini, Davide Delpiano, Armando Falcucci, Carla Figus, Federico Lugli, Gregorio Oxilia, Matteo Romandini. English proof-reading: Julien Riel-Salvatore.

## Acknowledgements

This project has been realised through funding from the European Research Council (ERC) under the European Union's Horizon 2020 research and innovation programme (grant agreement No 724046); http://www.erc-success.eu/. A thank should go to Stefano Ricci for helping in editing Figs. 7–9. Likewise we would like to warmly acknowledge the help of Birgitte Hoiberg Nielsen and Geoff Phillips for proof-reading the English text. Special thanks are due to Professors Arturo Palma di Cesnola and Paolo Gambassini for giving us the possibility of studying the material from their excavations. We would especially like to thank the Soprintendenze Archeologia, Belle Arti e Paesaggio per le Province di: Brindisi, Lecce e Taranto; Salerno e Avellino; città metropolitana di Genova; Province di Imperia, Savona e La Spezia for supporting our research and fieldwork over the years. We would also like to acknowledge the contribution from the Municipalities of Camerota, Castelcivita, Ginosa and Monte Argentario, and Parco Nazionale del Cilento e Vallo di Diano in the form of logistic support. The authors also thank the the Polo Museale della Liguria for facilitating and supporting fieldwork in Liguria. Recent fieldwork at Bombrini was funded also by the Université de Montréal, the Università di Genova, the Fonds Québécois pour la Recherche – Société et Culture (grant 2016-NP-193048), the Social Sciences and Humanities Research Council of Canada (Insight Grant 435-2017-1520) through grants to J. Riel-Salvatore and F. Negrino. Research at Fumane is coordinated by the Ferrara University (M.P.) in the framework of a project supported by MIBAC, public institutions (Lessinia Mountain Community, Fumane Municipality and others). Research at Riparo del Broion and Grotta di San Bernardino is designed by Ferrara University (M.P.) and was supported by MIBAC, the Province of Vicenza, the Veneto Region – Department for Cultural Heritage, and the Italian Ministry of Research and Education. Finally we would like to sincerely acknowledge the three anonymous reviewers who really enriched the quality of this paper thank to their valuable, appropriete and interesting comments related to the content, the philosophical background as well as the data presented in this paper.

## Appendix A. Supplementary data

Supplementary data to this article can be found online at https://doi.org/10.1016/j.quaint.2019.11.005.

## Uncited References